\documentclass[a4paper,11pt]{article}
\pdfoutput=1 

\usepackage{jheppub} 

\usepackage[T1]{fontenc} 

\usepackage{amsmath}
\usepackage{amsfonts}
\usepackage{fancyhdr}
\usepackage{hyperref}
\usepackage{listings}
\usepackage{inputenc}
\usepackage{xspace}
\usepackage{slashed}

\makeatletter
\def\@fpheader{\relax}
\makeatother

\newcommand{\beq}{\begin{equation}}
\newcommand{\eeq}[1]{\label{#1}\end{equation}}
\def\beqa{\begin{eqnarray}}
\def\eeqa#1{\label{#1}\end{eqnarray}}
\newcommand{\eeqn}{\end{equation}}

\def\bsp#1\esp{\begin{split}#1\end{split}}

\def\bem{\begin{enumerate}}
\def\eem{\end{enumerate}}
\def\btabt{\begin{tabular}{ll}}
\def\etabt{\end{tabular}}

\newcommand{\pythia}{{\sc Pythia}}

\newcommand{\delphes}{{\sc Delphes}}
\newcommand{\pgs}{{\sc PGS}}
\newcommand{\madgraph}{{\sc MadGraph}}

\newcommand{\python}{{\sc Python}}

\newcommand{\aloha}{{\sc ALOHA}}

\newcommand{\mrm}[1]{\mathrm{#1}}

\renewcommand{\b}{{\mathrm b}}
\renewcommand{\c}{{\mathrm c}}
\renewcommand{\d}{{\mathrm d}}
\newcommand{\e}{{\mathrm e}}

\newcommand{\g}{{\mathrm g}}

\newcommand{\n}{{\mathrm n}}
\newcommand{\p}{{\mathrm p}}

\newcommand{\s}{{\mathrm s}}
\renewcommand{\t}{{\mathrm t}}
\renewcommand{\u}{{\mathrm u}}

\newcommand{\D}{{\mathrm D}}

\renewcommand{\H}{{\mathrm H}}

\newcommand{\K}{{\mathrm K}}

\newcommand{\W}{{\mathrm W}}
\newcommand{\Z}{{\mathrm Z}}

\newenvironment{Itemize}{\begin{list}{$\bullet$}%
{\setlength{\topsep}{0.4mm}\setlength{\partopsep}{0.4mm}%
\setlength{\itemsep}{0.4mm}\setlength{\parsep}{0.4mm}}}%
{\end{list}}
\newcounter{enumct}
\newenvironment{Enumerate}{\begin{list}{\arabic{enumct}.}%
{\usecounter{enumct}\setlength{\topsep}{0.4mm}%
\setlength{\partopsep}{0.4mm}\setlength{\itemsep}{0.4mm}%
\setlength{\parsep}{0.4mm}}}{\end{list}}

\def\gev{~{\rm GeV}}

\def\lsim{\mathrel{\raise.3ex\hbox{$<$\kern-.75em\lower1ex\hbox{$\sim$}}}}
\def\gsim{\mathrel{\raise.3ex\hbox{$>$\kern-.75em\lower1ex\hbox{$\sim$}}}}
\def\ifmath#1{\relax\ifmmode #1\else $#1$\fi}

\newcommand{\Herwigpp}{{\sc Herwig++}\xspace}

\title{\boldmath From Lagrangians to Events: 
Computer Tutorial at the MC4BSM-2012 Workshop}

\author[a]{Stefan Ask,}
\author[b]{Neil D.~Christensen,}
\author[c]{Claude Duhr,}
\author[d]{Christophe Grojean,}
\author[e]{Stefan Hoeche,}
\author[f]{Konstantin Matchev,}  
\author[g,h]{Olivier Mattelaer,}
\author[i]{Stephen Mrenna,}
\author[j]{Andreas Papaefstathiou,}
\author[d]{Myeonghun Park,}
\author[k]{ Maxim Perelstein,}
\author[d]{and Peter Skands}

\affiliation[a]{Cavendish Laboratory, University of Cambridge, Cambridge CB3 0HE, UK.}
\affiliation[b]{PITTsburgh Particle physics, Astrophysics and Cosmology Center, Department of Physics \& Astronomy, University of Pittsburgh, Pittsburgh, PA 15260, USA.}
\affiliation[c]{Institute for Theoretical Physics, ETH Zurich, 8093 Zurich, Switzerland.}
\affiliation[d]{CERN Physics Department, Theory Division, CH-1211 Geneva 23, Switzerland.}
\affiliation[e]{SLAC, Stanford University, 2575 Sand Hill Rd., Menlo Park, CA 94025, USA.}
\affiliation[f]{Physics Department, University of Florida, Gainesville, FL 32611, USA.}
\affiliation[g]{Institut de Physique Th\'{e}orique and Centre for Particle Physics and Phenomenology (CP3),\\ Universit\'{e} Catholique de Louvain, B-1348 Louvain-la-Neuve, Belgium.}
\affiliation[h]{Department of Physics, University of Illinois at Urbana-Champaign, Urbana, IL 61801.}
\affiliation[i]{SSE Group, Computing Division, Fermilab, Batavia, IL 60510, USA.}
\affiliation[j]{Institut fu\"{r} Theoretische Physik, Universit\"{a}t Z\"{u}rich, CH-8057 Z\"{u}rich, Switzerland.}
\affiliation[k]{Department of Physics, Cornell University, Ithaca, NY 14853 USA. }

\emailAdd{stefan.ask@cern.ch}
\emailAdd{neilc@pitt.edu}
\emailAdd{duhrc@itp.phys.ethz.ch}
\emailAdd{christophe.grojean@cern.ch}
\emailAdd{shoeche@slac.stanford.edu}
\emailAdd{matchev@phys.ufl.edu}
\emailAdd{olivier.mattelaer@uclouvain.be}
\emailAdd{mrenna@fnal.gov}
\emailAdd{andreasp@physik.uzh.ch}
\emailAdd{myeonghun.park@cern.ch}
\emailAdd{mp325@cornell.edu}
\emailAdd{peter.skands@cern.ch}

\abstract{This is a written account of the computer tutorial offered at the Sixth MC4BSM workshop at
Cornell University, March 22-24, 2012. The tools covered during the tutorial include: {\sc FeynRules}, 
{\sc LanHEP}, {\sc MadGraph}, {\sc CalcHEP}, {\sc Pythia 8}, {\sc Herwig++}, and {\sc Sherpa}. 
In the tutorial, we specify a simple extension of the Standard Model, at the level of a Lagrangian. 
The software tools are then used to automatically generate a set of Feynman rules, 
compute the invariant matrix element for a sample process, and generate both 
parton-level and fully hadronized/showered Monte Carlo event samples. 
The tutorial is designed to be self-paced, and detailed instructions for all steps are included in 
this write-up. Installation instructions for each tool on a variety of popular platforms are also provided.}

\begin{document} 
\maketitle
\flushbottom

\section{Introduction}
\label{sec:introduction}


With the advent of the LHC, the field of particle physics has entered an exciting new era. 
A direct exploration of the TeV energy scale has finally begun. A large number of 
extensions of the Standard Model (SM), or ``Beyond-the-Standard" Models (BSMs) 
for short, have been proposed over the years, and many of them can be probed by the LHC experiments. 
To make such tests possible, theoretical predictions of each model must be computed
at a level that allows direct comparison to data. Monte Carlo (MC) generators are the 
basic tool for obtaining such predictions.

\begin{figure}[t]
\centering
\includegraphics[width=5.0in]{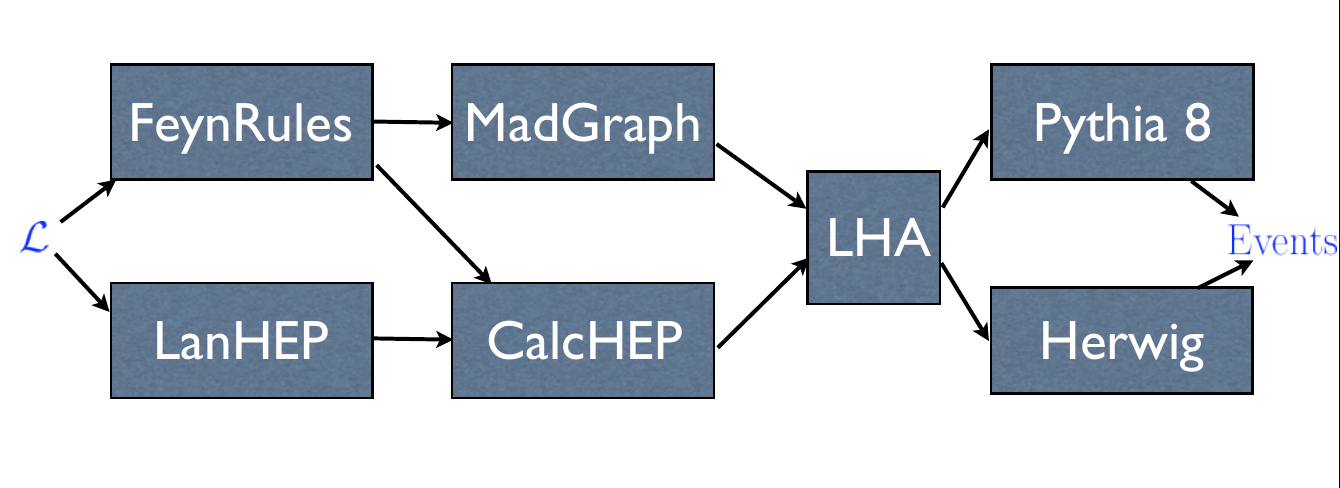}
\caption{Simulation path from the Lagrangian ${\cal L}$ of a BSM to a particle-level Monte 
Carlo event sample. All tools shown here are included in this tutorial.}
\label{fig:path}
\end{figure} 

Any BSM consistent with basic postulates of quantum mechanics and special relativity 
is a quantum field theory (QFT) defined by specifying its field content and the interaction 
Lagrangian ${\cal L}$. As long as all interactions can be treated perturbatively, the fully 
differential cross section for any scattering process of interest can be computed by following 
the standard procedure of QFT perturbation theory. Namely, starting from ${\cal L}$, one derives a 
set of Feynman rules, draws all possible Feynman diagrams for the process of interest, 
writes down the invariant matrix element ${\cal M}$, squares it, and multiples it by the phase 
space weight. Since this procedure is completely unambiguous, it should be possible to 
teach a computer to perform it automatically. A set of automated tools to achieve this task 
has recently been developed, see Fig.~\ref{fig:path}. {\sc FeynRules}~\cite{Christensen:2008py} 
and {\sc LanHEP}~\cite{Semenov:2008jy} 
perform the derivation of a complete set of Feynman rules from a given Lagrangian. 
The output of these programs can then be communicated to an automated matrix element 
calculator such as {\sc MadGraph}~\cite{Stelzer:1994ta} or {\sc CalcHEP}~\cite{Belyaev:2012qa}
(or its sister package {\sc CompHEP} \cite{Pukhov:1999gg,Boos:2009un}) which generates 
$|{\cal M}|^2$ for any process specified by the user. This object in turns serves as an 
input to the Monte Carlo simulation of the process, first at the parton level with parton-level 
Monte Carlo event generators like {\sc MadGraph} or {\sc CalcHEP} ({\sc CompHEP}). 
They create a set of ``events'', where each event 
is a record containing the identity and four-momentum of each of the initial- and 
final-state particles in the hard scattering process. 
The events are distributed in momenta, helicities, {\it etc.} according to the invariant matrix 
element $|{\cal M}|^2$ computed in the previous step. In effect, the MC generator performs 
a pseudo-experiment: the generated set of events is just a particular statistical realization 
of the distributions predicted by the theory for the process of interest. 
The produced parton-level events are in turn handed to a general purpose event generator 
such as {\sc Pythia}~\cite{Sjostrand:2007gs} or {\sc Herwig}~\cite{Marchesini:1991ch,Corcella:2000bw}, 
which creates complete events, 
including the effects of fragmentation and hadronization of colored particles, initial and final 
state radiation via parton showers, effects from the underlying event, decays of unstable resonances, {\it etc.}
The communication between the two classes of generators (parton-level and general purpose)
is done following a universally accepted ``Les Houches Accord" (LHA) format  \cite{Boos:2001cv,Alwall:2006yp}.

The program of the Sixth MC4BSM workshop held at
Cornell University in March 2012 centered around hands-on computer tutorials \cite{MC4BSM} 
illustrating the tools depicted in Fig.~\ref{fig:path}. 
The starting point of the tutorials was a simple toy theory model
described in Section \ref{sec:model}. Its Feynman rules can then be automatically derived
via either {\sc FeynRules} (Section \ref{sec:FR}) or {\sc LanHEP} (Section \ref{sec:LH}), 
the end product being the input files necessary to define the model in the parton-level event 
generators {\sc MadGraph} and {\sc CalcHEP}. The next step is to produce parton-level 
events, which is done in the tutorials covered in Sections~\ref{sec:MG}
and \ref{sec:CH}, respectively. The parton-level events are then fed into
{\sc Pythia 8} (Section \ref{sec:PY}) or \Herwigpp (Section \ref{sec:HW}).
Finally, Section \ref{sec:SH} contains a special tutorial on {\sc Sherpa} \cite{Gleisberg:2003xi},
which provides an alternative path combining all of the above steps.
In principle, each tutorial exercise is self-contained and independent 
of the others. Typically, each tutorial has two parts: in the first part (the pre-workshop exercise)
students download and install the software and perform some simple tests to make 
sure it runs properly, while the second part deals with the actual physics simulations.

\section{The reference BSM model used in the tutorials}
\label{sec:model}

The tutorial exercises are illustrated with a toy reference BSM model whose particle 
content is shown in Table~\ref{tab:fields}.
\begin{table}[tbp]
\centering
\begin{tabular}{|c|c|c|c|c|c|}
\hline
Notation &      Spin & Mass     &  SU(3)   &  SU(2)   & U(1) \\
\hline 
$\Phi_1$ &           0 & $M_1$   &      1       &      1      &     0  \\
$\Phi_2$ &           0 & $M_2$    &      1      &      1       &     0 \\
$U$        &       1/2  & $M_U$    &    3        &      1      &     2/3 \\
$E$        &      1/2   & $M_E$     &    1       &      1       &    -1 \\
\hline
\end{tabular}
\caption{\label{tab:fields} The BSM field content (with quantum numbers) of the reference toy model.}
\end{table}
The model contains two real scalar fields, $\phi^1$ and $\phi^2$. They are singlets under all SM gauge groups. 
Their mass terms are\footnote{All Lagrangian parameters, here and below, are assumed to be real.}:
\beq
{\cal L}_{\rm s.m.}=-\frac{m_1^2}{2} \phi_1^2- \frac{m_2^2}{2} \phi_2^2-m_{12}^2\phi_1\phi_2\,.
\eeq{Phi_masses} 
The corresponding mass eigenstates will be denoted by $\Phi_1$ and $\Phi_2$, and their mass eigenvalues 
by $M_1$ and $M_2$, respectively. For definiteness we will assume that $M_1<M_2$.

The model also contains two new Dirac fermion fields, $U$ and $E$. Their SM quantum numbers 
are those of the SM $u_R$ and $e_R$, respectively. These fields have mass terms
\beq
{\cal L}_{\rm f.m.}= M_U \bar{U} U + M_E \bar{E} E\,.
\eeq{Fmasses}
and interact with the new scalars via
\beq
{\cal L}_{\rm Yuk} = \lambda_1  \phi_1 \bar{U} P_R u+ \lambda_2 \phi_2\bar{U} P_R u  + \lambda^\prime_1  \phi_1\bar{E} P_R e + \lambda^\prime_2 \phi_2\bar{E} P_R e \,,
\eeq{Yuks}
where $u$ and $e$ are the SM up-quark and electron fields.
Note that there is a ${\cal Z}_2$ symmetry under which all fields we added ($\phi_{1,2}$, $U$, $E$) 
flip sign, while all SM fields do not, so the new particles must be pair-produced, and the lightest 
new particle (LNP) is stable. This same ${\cal Z}_2$ also forbids $U-u$ and $E-e$ mixing via 
Yukawa couplings with the SM Higgs. 

We assume the following ordering of masses:
\beq
M_U > M_2 > M_L > M_1\,,
\eeq{massorder}
so that $\Phi_1$ is the LNP. Not having any SM interactions, it appears as MET in the detector. 
The ultimate goal of the tutorial is to simulate the process
\beq
pp \to \bar{U}U\,,
\eeq{prod}
at an 8 TeV LHC, and the subsequent $U$ decays:
\beqa
U &\to& u \Phi_1 \,,\\
U &\to& u \Phi_2\,,~~\Phi_2\to e E\,,~~E \to e \Phi_1 \,.
\eeqa{decays}

\section{{\sc FeynRules} tutorial}
\label{sec:FR}

The {\sc FeynRules} tutorial consists of two steps: a pre-workshop exercise described in
Section \ref{sec:FRpre}, and the on-site exercise described in Section~\ref{sec:FRonsite}

\subsection{Installation instructions}
\label{sec:FRpre}

The aim of this session is to make the user familiar with the basics of {\sc FeynRules}
using the example of the Standard Model included in the distribution.
{\sc FeynRules} can be downloaded from {\tt http://feynrules.irmp.ucl.ac.be}. You need to have Mathematica installed on your machine in order to run {\sc FeynRules}. Simply download the package and untar it.

Next, you can have a look at the SM implementation, contained in the directory \verb+/feynrules-current/models/SM/+. This folder contains various files, but in this introductory session we will only look at two of them:
\begin{enumerate}
\item \verb+SM.fr+: the model file (a text file) containing the implementation of the SM in {\sc FeynRules}.
\item \verb+SM.nb+: a Mathematica notebook, showing how to load and run the model to obtain the Feynman rules.
\end{enumerate}

\subsubsection{The {\sc FeynRules} model file}
Open the model file for the SM, \verb+SM.fr+, in a text editor.
In the following we give a very brief account on the general structure of a {\sc FeynRules} model file.
During the tutorial, you will be given the opportunity to write your own {\sc FeynRules} model file, and we therefore limit ourselves at this stage to describe the generic features.

A generic {\sc FeynRules} model file contains four parts:
\begin{enumerate}
\item A list called \verb+M$GaugeGroups+ containing the definitions of all the gauge groups of the model. 
\item A list called \verb+M$Parameters+ containing the definitions of all the parameters of the model. 
\item A list called \verb+M$ClassesDeclarations+ containing the definitions of all the fields of the model. Note that the fields are grouped into classes of particles carrying the same quantum numbers.
\item The Lagrangian of the model, written in Mathematica form.
\end{enumerate}

\subsubsection{The sample notebook}
Next, open the notebook \verb+SM.nb+. The first few lines of the notebook show how to load {\sc FeynRules} into Mathematica,
\begin{verbatim}
$FeynRulesPath = SetDirectory["< address of the package >"]

<< FeynRules`

SetDirectory[$FeynRulesPath <> "/Models/SM"];
\end{verbatim}
The first line sets the path to the {\sc FeynRules} main directory. Note that \verb+< address of the package >+ should be replaced by the path of the {\sc FeynRules} main directory on your computer. For example, if the {\sc FeynRules} main directory is in your home folder, then the first line should read
\begin{verbatim}
$FeynRulesPath = SetDirectory["~/feynrules-current"]
\end{verbatim}
The second line loads the {\sc FeynRules} package. If successful, an output is printed on the screen. The third line changes the path to the subfolder \verb+Models/SM+ that contains the SM implementation.

After {\sc FeynRules} has been successfully loaded into Mathematica, the SM implementation can be loaded by issuing the command
\begin{verbatim}
LoadModel["SM.fr"]
\end{verbatim}
The next command is optional, and loads some additional files that put the masses of the first two generations of fermions to zero, and restricts the CKM matrix to be diagonal. As this step is optional, we will not describe it any further.

The first section of the notebook, \emph{The SM Lagrangian}, shows the Lagrangian that is implemented in \verb+SM.fr+. Note that the SM implementation of {\sc FeynRules} is available in both Feynman and unitary gauge. The boolean variable \verb+FeynmanGauge+ allows to switch between the two gauges.
The second section shows how one can use some built-in functions to perform some basic checks on the Lagrangian, \emph{e.g.}, to check if the Lagrangian is hermitian. The last two sections illustrate how to compute the Feynman rules of the model with {\sc FeynRules}, and how to use the interfaces to the Feynman diagram generators.

\subsection{On-site tutorial}
\label{sec:FRonsite}

The aim of this tutorial is to implement a simple extension of the SM into {\sc FeynRules}.
The model itself is described in Section~\ref{sec:model}. In the following we give detailed 
instructions of how to implement this model into {\sc FeynRules} and how to obtain the 
corresponding model files for {\sc CalcHEP}, {\sc MadGraph 5} and {\sc Sherpa}.

\subsubsection{Preparation of the model file}
As the model we are going to implement is a simple extension of the SM, it is not necessary to start from scratch, but we can use the implementation of the SM included in the folder \verb+/Models/SM+. We therefore start by making a copy of this folder, in order to keep a clean version of the SM. To do so, change directory to the \verb+Models+ subdirectory and make a copy of the SM folder, before going into the new directory
\begin{verbatim}
cd Models
cp -r SM MC4BSM
cd MC4BSM
\end{verbatim}
Even though the implementation is based on the model file \verb+SM.fr+ for the Standard Model, the SM sector of the model will be implemented into a separate file that will simply be loaded on top of \verb+SM.fr+. We therefore start by opening a blank text file called \verb+MC4BSM.fr+.
You can start by personalizing the model file by including a name for you model, the name of the author, \emph{etc.},
\begin{verbatim}
M$ModelName = "MC4BSM_2012";

M$Information = {Authors      -> {"C. Duhr"}, 
                 Version      -> "1.0",
                 Date         -> "27. 02. 2012",
                 Institutions -> {"ETH Zurich"},
                 Emails       -> {"duhrc@itp.phys.ethz.ch"}
             };
\end{verbatim}
Note that his information is purely optional and could be omitted.

\subsubsection{Implementation of the new parameters}
We start by implementing the new parameters. The model we are considering depends on 9 new parameters, which we assume to be real in the following. 
{\sc FeynRules} distinguishes between two types of parameters, the so-called \emph{external} parameters given as numerical inputs and the \emph{internal} parameters, related to other external and/or internal parameters via algebraic expressions. All the parameters of the model, both external and internal, are given in the {\sc FeynRules} model file as a list named \verb+M$Parameters+. Note that if an internal parameter $x$ depends on some other parameter $y$, then $y$ should appear in  \verb+M$Parameters+ \emph{before} the internal parameter $x$.

The new external parameters of the model are
\begin{itemize}
\item 5 mass parameters: $m_1$, $m_2$, $m_{12}$, $M_U$, $M_E$.
\item 4 coupling constants: $\lambda_1$, $\lambda_2$, $\lambda'_1$, $\lambda'_2$.
\end{itemize}
Note however that there is a difference between the mass parameters in the scalar and fermonic sectors: while the masses in fermonic sector are physical masses, the mass matrix in the scalar sector is not diagonal. For this reason, we will not discuss in the following the fermion masses $M_U$ and $M_E$, as they will be defined together with the particles rather than as parameters of the model.

Let us now turn to the definition of the mass parameters in the scalar sector. The masses $m_1$, $m_2$ and $m_{12}$ will be denoted in the {\sc FeynRules} model file by \verb+MM1+, \verb+MM2+ and \verb+MM12+. In the following we only show how to implement \verb+MM1+, all other cases being similar. \verb+MM1+ corresponds to the following entry in the list \verb+M$Parameters+,
\begin{verbatim}
M$Parameters = {
    ...
    MM1 == {
           ParameterType -> External,
           Value         -> 200
         },
    ...
}
\end{verbatim}
The first option tags \verb+MM1+ as an external parameter, while the second option assign a value of 200GeV to $m_1$. We stress that this numerical value can be changed later on in the matrix element generators.

The masses in the scalar sector are not the physical masses, because the mass matrix is not diagonal.
In order to obtain the physical masses, we need to diagonalize the mass matrix
\begin{equation}
\left(\begin{array}{cc}
m_1^2 & m_{12}^2 \\
m_{12}^2 & m_2^2
\end{array}\right)\,.
\end{equation}
In the following, we denote the eigenvalues by \verb+MPe1+ and \verb+MPe2+. In addition, we need to introduce a mixing angle $\theta$ (\verb+th+) relating the fields $\phi_i$ to the mass eigenstates $\Phi_i$ by,
\begin{equation}
\left(\begin{array}{c}
\phi_1 \\
\phi_2
\end{array}\right) = \left(\begin{array}{cc}
-\sin\theta & \cos\theta \\
\cos\theta & \sin\theta
\end{array}\right)\,\left(\begin{array}{c}
\Phi_1 \\
\Phi_2
\end{array}\right)\,.
\end{equation}
As in this case the mass matrix is only two-dimensional, we can compute the eigenvalues and the mixing angle analytically, and simply implement the analytical formulas into {\sc FeynRules}. The implementation follows exactly the same lines as for the masses $m_1$, $m_2$, $m_{12}$, with the only differences that
\begin{enumerate}
\item the \verb+ParameterType+ is \verb+Internal+ (as these parameters are dependent on the external mass parameters,
\item the \verb+Value+ is given by an analytical expression (in Mathematica syntax).
\end{enumerate}

Next we turn to the implementation of the new coupling constants, which we will call \verb+lam1+, \verb+la21+, \verb+lam1p+, \verb+lam2p+. They are all external parameters, and thus the implementation follows exactly the same lines as the implementation of the mass parameters, with only one modification: some matrix element generators, like for example {\sc MadGraph}, keep track of the types of couplings that enter a process. This allows for example to generate a process by only taking into account QCD-type vertices, and to neglect all QED-type vertices. For this reason, it is mandatory to tell the matrix element generator how the new coupling constants should be counted. As in this case we are dealing with new classes of couplings which are \emph{a priori} independent of QCD or QED interactions, we simply assign a new tag, called \emph{interaction order}, to the coupling via the option
\begin{verbatim}
InteractionOrder ->  {NP, 1}
\end{verbatim}
The name of the tag (\verb+NP+ for ``new physics'' in this case) can be chosen freely. The above option instructs the matrix element generator to count one unit of ``\verb+NP+'' for each new coupling.

\subsubsection{Implementation of the fields}
In this section we discuss the implementation of the new fields. The implementation is similar to the implementation of the parameters, \emph{i.e.}, all the fields are entries of a list called \verb+M$ClassesDescription+. In Tab.~\ref{tab:particles} we show the names of the fields used in the implementation\footnote{Note that the symbol {\tt u}, {\tt e} and {\tt phi} are already in use in the SM implementation. We also avoid using simply uppercase letters, as some matrix element generators are case insensitive.}.
\begin{table}[!h]
\begin{center}
\begin{tabular}{|c|c|c|c|c|c|}
\hline
$U$ & $E$ & $\phi_1$& $\phi_2$ & $\Phi_1$& $\Phi_2$ \\
\hline
\verb+uv+ & \verb+ev+ & \verb+pi1+ & \verb+pi2+ & \verb+p1+ &  \verb+p2+\\
\hline
\end{tabular}
\caption{\label{tab:particles}Symbols used for the fields in the {\sc FeynRules} implementation.}
\end{center}
\end{table}

We illustrate the implementation of a new field on the example of the particle $U$ ({\tt uv}). The definition of the particle corresponds to an entry in \verb+M$ClassesDescription+ of the following form
\begin{verbatim}
M$ClassesDescription = {
      ...
      F[10] == {
          ClassName      -> uv,
          SelfConjugate  -> False,
          Indices        -> {Index[Colour]},
          QuantumNumbers -> {Y -> 2/3, Q -> 2/3},
          Mass           -> {Muv, 500},
          Width          -> {Wuv,1}
         },
       ...
 }
 \end{verbatim}
The meaning of this definition is as follows: each particle class has a name of the form \verb+X[i]+, where \verb+X+ is related to the spin of the field (See Tab.~\ref{tab:classes}), and {\tt i} is an integer that labels the classes. Note that {\tt i} can be chosen freely, as long as there is no name clash with an already existing class (in this case, there could be a name clash with the SM particles already defined in {\tt SM.fr}). Each class has a series of options 
\begin{enumerate}
\item \verb+ClassName+: the symbol by which the particle will be represented in the Lagrangian.
\item \verb+SelfConjugate+: a boolean variable, indicating whether the particle has an antiparticle (\verb+False+) or not (\verb+True+). If the field is not self-conjugate, a symbol for the antiparticle is automatically defined by appending ``{\tt bar}'' to the name of the particle. In the above example the antiparticle associated to {\tt uv} will be denoted by {\tt uvbar}. Note that in the case of fermions the symbol for the antiparticle refers to the quantity $\bar{U}$ rather than $U^\dagger$.
\item {\tt Indices}: All indices carried by the field. The available types of indices from the SM implementation are
\begin{itemize}
\item {\tt Generation}: fermion flavor index ranging from 1 to 3,
\item {\tt Colour}: fundamental color index ranging from 1 to 3,
\item {\tt Gluon}: adjoint color index ranging from 1 to 8,
\item {\tt SU2W}: adjoint $SU(2)_L$ index ranging from 1 to 3.
\end{itemize}
\item {\tt QuantumNumbers}: a list of all $U(1)$ charges carried by the field. In the SM implementation the following $U(1)$ charges are already defined
\begin{itemize}
\item {\tt Y}: weak hypercharge,
\item {\tt Q}: electric charge.
\end{itemize}
\item {\tt Mass}: the mass of the particle. It is a list of two elements, the first being the symbol used to represent the mass, and the second its value (in GeV). If the value of the mass is obtained from some analytic expression defined as an internal parameter with the same symbol (as is the case for example in the scalar sector of the model), the value is set to {\tt Internal}.
\item {\tt Width}: the width of the particle. The definition is similar to {\tt Mass}. Note that as we do not yet know the widths of the new particles, we simply set it for now to 1GeV, and will determine its exact value later using one of the matrix element generators.
\end{enumerate}
The implementation of the other mass eigenstates (\verb+ev+, \verb+p1+, \verb+p2+) is similar, so we do not discuss it here. 
\begin{table}[!ht]
\begin{center}
\begin{tabular}{|c|ccccc|}
\hline
Spin & 0 & 1/2 & 1 & 2 & ghost \\
\hline
Symbol & {\tt S} &  {\tt F} &  {\tt V} &  {\tt T} &  {\tt U}\\
\hline
\end{tabular}
\caption{\label{tab:classes} Available particle classes in {\sc FeynRules}.}
\end{center}
\end{table}

Let us comment on the implementation of the interaction eigenstates $\phi_i$. Indeed, while the matrix element generators work exclusively at the level of the mass eigenstates, the interaction eigenstates are in general useful to write the Lagrangian in a compact form. It is therefore useful to define also the fields for the interaction eigenstates $\phi_i$. The definition of these fields is similar to the mass eigenstates, \emph{e.g.},
\begin{verbatim}
    S[10] == {
          ClassName     -> pi1,
          SelfConjugate -> True,
          Indices       -> {},
          Unphysical    -> True,
          Definitions   -> {pi1  ->  - Sin[th] p1 + Cos[th] p2}
         },
\end{verbatim}
First, note that the {\tt Mass} and {\tt Width} options are omitted\footnote{The {\tt QuantumNumbers} option is also omitted, but for the simple reason that the fields $\phi_i$ do not carry any $U(1)$ charges.}, as these fields are not mass eigenstates. This last fact is made explicit by the option
\begin{verbatim}
Unphysical    -> True,
\end{verbatim}
which instruct {\sc FeynRules} not to output this field to a matrix element generator. Finally, the relation of the field {\tt pi1} to the mass eigenstates is simply given as a Mathematica replacement rule in the {\tt Definitions} option.
          
\subsubsection{Implementation of the Lagrangian}
The definitions in the model file being complete, we now turn to the implementation of the Lagrangian. This can be done either in the model file, or directly in a Mathematica notebook. Here we use the latter approach, and we start by opening a new notebook and load the {\sc FeynRules} package (see the pre-installation instructions). Next we have to load the model files, both for the SM and for the new sector,
\begin{verbatim}
LoadModel["SM.fr", "MC4BSM.fr"]
\end{verbatim}
Note that the new model file should be loaded after {\tt SM.fr}. Furthermore, we also load two additional files, which restrict the first two fermion generations to be massless and the CKM matrix to be diagonal,
\begin{verbatim}
LoadRestriction["DiagonalCKM.rst", "Massless.rst"]
\end{verbatim}

The new Lagrangian consists of three parts,
\begin{equation}
\begin{cal}L\end{cal} = \begin{cal}L\end{cal}_{scalar, kin} + \begin{cal}L\end{cal}_{fermion, kin} + \begin{cal}L\end{cal}_{Yuk}\,.
\end{equation}
The kinetic terms for the new scalars can easily be implemented by using the symbols for the gauge eigenstates and the mass parameters defined in the model file, as well as the symbol for the space-time derivative $\partial_\mu$ in {\sc FeynRules}, {\tt del[ ..., mu]}. As an example, we have
\begin{center}
\begin{tabular}{c}
$\displaystyle{1\over 2}\partial_\mu\phi_1 + \partial^\mu\phi_1 -{1\over 2}m_1^2\phi_1^2$ \\
\verb+1/2 del[p1, mu] del[p1, mu] - 1/2 MM1^2 p1^2+
\end{tabular}
\end{center}
The kinetic terms for the fermions can be implemented in a  similar way. However, as the fermions are charged under the SM gauge group, we have to use the covariant derivative {\tt DC} rather than the space-time derivative {\tt del}. Furthermore, we have to use a ``.'' instead of an ordinary multiplication in order to take the non-commuting nature of the fermions into account. As an example, we have
\begin{center}
\begin{tabular}{c}
$\displaystyle i\,\bar{U}\gamma^\mu D_\mu U -M_U \bar{U}U$ \\
\verb+ I uvbar.Ga[mu].DC[uv, mu] - Muv uvbar.uv+
\end{tabular}
\end{center}
where {\tt Ga[mu]} is the {\sc FeynRules} symbol for the Dirac matrix $\gamma^\mu$.
Finally, the Yukawa interactions can be implemented in the same way as the kinetic terms for the fermions, \emph{e.g.},
\begin{center}
\begin{tabular}{c}
$\displaystyle \lambda_1\,\phi_1\,\bar{U}P_+u$ \\
\verb+lam1 pi1 uvbar.ProjP.u+
\end{tabular}
\end{center}
where {\tt u} denotes the $u$ quark field defined in {\tt SM.fr} and {\tt ProjP} denotes the right chiral projector (the left projector is denoted by {\tt ProjM}). Note that {\sc FeynRules} contains a function \verb+HC[ ]+ which allows to obtain the hermitian conjugate of an expression in an automated way.

\subsubsection{Computing the Feynman rules and running the interfaces}
Our model implementation is now complete, and so we can compute the Feynman rules.
The Feynman rules of the new sector can be obtained by issuing the command
\begin{verbatim}
FeynmanRules[ LNew ]
\end{verbatim}
where \verb+LNew+ is the name of the variable that contains the new Lagrangian.

The Feynman rules can be written to file in a format suitable to various matrix element generators by using the {\sc FeynRules} interfaces. In this tutorial, we will use the interfaces to {\sc CalcHEP}, {\sc MadGraph 5} and {\sc Sherpa}, which can be called via
\begin{verbatim}
WriteCHOutput[ LSM + LNew ];
WriteUFO[ LSM + LNew ];
WriteSHOutput[ LSM + LNew ];
\end{verbatim}
where {\tt LSM} is the SM Lagrangian implemented in {\tt SM.fr}.
Note that the SM implementation is available in both Feynman gauge and unitary gauge. A boolean variable {\tt FeynmanGauge} allows to switch between both gauges. While {\sc CalcHEP} and {\sc MadGraph 5} support both gauges, {\sc Sherpa} only supports unitary gauge, and the {\tt FeynmanGauge} variable should correspondingly be set to {\tt False}.

\section{{\sc LanHEP} tutorial}
\label{sec:LH}

 
{\sc LanHEP} is a program which allows to create input model files for the programs {\sc CompHEP} \cite{comphep}
or {\sc CalcHEP} \cite{calchep} from a Lagrangian provided by the user. 
This program is as easy as \LaTeX.  We define the symmetries, the particles and then
write down a Lagrangian, just as we would write them in a paper. 
Then we compile the write-up  and obtain {\sc CompHEP}- or {\sc CalcHEP}-specific model files for further collider studies.
In short, if you can write down a Lagrangian, and if you have enough knowledge 
of \LaTeX, then your skills are probably enough to use {\sc LanHEP} as well.

An excellent reference for beginners is the {\sc LanHEP} manual \cite{Semenov:2010qt}. 
Here we will demonstrate how to use {\sc LanHEP} with the specific example (see Section~\ref{sec:model}) 
provided for the MC4BSM 2012 workshop\footnote{MC4BSM 2012 provides a nice homepage \cite{MC4BSM} for all tutorials.}.
First let us discuss how to install {\sc LanHEP}.

\subsection{Installation}
Installation is never an easy job, especially when there are several choices for the operating system.  
But if you have a working C-compiler in your operating system, there should be no problem. 
\begin{itemize}
\item[1] Linux with gcc.
\item[2] OS/X: to use gcc, you need to install Xcode in Apple Appstore. But to use Monte Carlo simulations, 
you will also need a fortran compiler.  Free fortran compilers are available from these sites \cite{OSX1, OSX2}.\footnote{Starting from Mountain Lion, in Xcode, you need to install ``Command Line Tools''
which you can find in ``Xcode/Preferences/Downloads". The current version coming from Xcode is gcc 4.2, 
thus we recommend that you match your versions of gcc, g++ and gcov with the version of gfortran that you download from a third party.}
\item[3] Windows: we recommend that you use a virtual machine\footnote{There is a free virtual machine
software called VirtualBox\,\cite{virtualbox}. } to install Linux under the Windows system 
or have a dual boot with a Linux partition.
Most Monte Carlo simulations are too complicated to run under Cygwin.
\end{itemize}

The detailed installation instructions are well explained on the {\sc LanHEP} webpage \cite{lanhep}. 
Here we will briefly recap the procedures based on the file, ``lanhep315.tar" which is the
current version at the time of this writing. 
\begin{description}
\item[1.] Download the {\sc LanHEP} program from the {\sc LanHEP} webpage \cite{lanhep}.
\item[2.] Make a proper directory for the {\sc LanHEP} installation. For example 
\begin{lstlisting}
mkdir $HOME/MC4BSM 
\end{lstlisting}
\item[3.] Move the downloaded file to the directory that you just created: 
\begin{itemize}
\item[a.] Let's suppose that you made an MC4BSM directory under your \$HOME directory, 
and you downloaded a ``lhep315.tar" file in the `\$HOME/Downloads" directory, 
then you can move the file into your MC4BSM folder by typing
\begin{lstlisting}
mv $HOME/Downloads/lhep315.tar $HOME/MC4BSM/.
\end{lstlisting}
\item[b.] Unzip ``lhep315.tar" file by typing 
\begin{lstlisting}
tar -xvf lhep315.tar
\end{lstlisting}
\end{itemize} 
\item[3.] Go to the ``lanhep315" directory and read the ``README" file. 
\item[4.] Compile {\sc LanHEP} by typing
\begin{lstlisting}
make
\end{lstlisting}
\end{description}

\subsection{Warm-Up exercise}

Before we begin the real MC4BSM exercise, it would be good to start with a 
very simple example, ``QED", which is the first example in the {\sc LanHEP} manual version 3.0 \cite{Semenov:2008jy}. 
You can find its model files in the ``mdl" directory as ``qed.mdl".  The contents of this file are the following
\begin{lstlisting}[frame=single]
model qed/1.
parameter ee = 0.3133: 'Electric charge'.
vector A/A:photon.
let F^mu^nu=deriv^mu*A^nu-deriv^nu*A^mu.
spinor e1:(electron, mass me=0.000511).
lterm ee*E1*gamma*A*e1.
\end{lstlisting}
Without looking at the manual, we can easily guess that there are two fields, 
an electron (e1) and a photon (A).  In this particular model file, there are missing 
terms from the Lagrangian (if you compare it to the text in the manual). 
You can modify this model file by writing down the missing terms.
Thus a complete simple ``QED" model file will look like

\begin{lstlisting}[frame=single]
model qed/1.
parameter ee=0.31333: 'elementary electric charge'. 
spinor e1/E1:(electron, mass me=0.000511).
vector A/A:(photon).
let F^mu^nu=deriv^mu*A^nu-deriv^nu*A^mu.
lterm -1/4*(F^mu^nu)**2 - 1/2*(deriv^mu*A^mu)**2. 
lterm E1*(i*gamma*deriv+me)*e1.
lterm ee*E1*gamma*A*e1.
\end{lstlisting}

At this stage we can ask some simple questions:
\begin{itemize}
\item[1.] What is the relation between e1 and E1?
\item[2.] How can we use a covariant derivative instead of splitting the Lagrangian into 
pieces?\footnote{See Section 3.4 in the manual 3.0 \cite{Semenov:2008jy}.}
\item[3.] What is the unit of mass in {\sc LanHEP}?
\end{itemize}

In order to create the model files for the actual {\sc CompHEP} or {\sc CalcHEP} program, we need to 
compile this model file. It would be better to make a separate directory for the "compiled" model. 
Let's make the directory inside your {\sc LanHEP} directory (if you are in the mdl directory, 
go one level up). For example,
\begin{itemize}
\item[] 
\begin{lstlisting}[language=sh]
mkdir QED
\end{lstlisting}
\end{itemize}
Now compile the ``qed.mdl" file by using the ``lhep" command as follows\footnote{If you type  ``./lhep" without any options,  you will see the available choices.}
\begin{itemize}
\item[]
\begin{lstlisting}[language=sh]
./lhep  -OutDir QED mdl/qed.mdl -CalcHEP 
\end{lstlisting}
\end{itemize}
You will get this ``warning'' message 
\begin{lstlisting}[language=sh,frame=single]
File mdl/qed.mdl processed, 0 sec.

Warning: property 'pdg' is not defined for particle 'e1'
\end{lstlisting}
``PDG" is a number which is defined for each particle according to the Particle Data Group for various Monte Carlo simulators. 
For example the PDG number for an electron is 11. Our particle ``e1" defined in the ``qed.mdl'' file 
was not assigned a PDG number. There are two ways to resolve this warning. 
If you open the ``calchep.rc" file in the mdl directory, you will see the following contents.

		   
\begin{lstlisting}[language=sh,frame=single]
prtcformat fullname: 'Full   Name     ', 
           name:' P ', aname:' aP', pdg:'  number  ',
           spin2,mass,width, color, aux, texname: ' LaTeX(A)       ',  
           atexname:'  LateX(A+)      '.

prtcproperty pdg:(A=22, Z=23, 'W+'=24, G=21, 
				d=1, u=2, s=3, c=4, b=5, t=6,
                ne=12, nm=14, nl=16,
                e=11, m=13, l=15,
  ~ne=1000012, ~nm=1000014, ~nl=1000016,  
  ~e1=1000011, ~m1=1000013, ~l1=1000015,
  ~e2=2000011, ~m2=2000013, ~l2=2000015,
  ~eL=1000011, ~mL=1000013,
  ~eR=2000011, ~mR=2000013,
  
  ~u1=1000002, ~c1=1000004, ~t1=1000006,
  ~u2=2000002, ~c2=2000004, ~t2=2000006,
  ~uL=1000002, ~cL=1000004,
  ~uR=2000002, ~cR=2000004,

  ~d1=1000001, ~s1=1000003, ~b1=1000005,
  ~d2=2000001, ~s2=2000003, ~b2=2000005,
  ~dL=1000001, ~sL=1000003,
  ~dR=2000001, ~sR=2000003,

   h=25, H=35, H3=36, 'H+'=37,
   ~o1=1000022, ~o2=1000023, ~o3=1000025, ~o4=1000035,
   '~1+'=1000024, '~2+'=1000037, ~g=1000021).
\end{lstlisting}
Thus, we can edit this file directly to include a PDG number (11) for our field `e1', or we can use 
a function ``prtcproperty pdg" during our write-up of the Lagrangian.

\subsubsection{Output}
After compiling the model file ``qed.mdl", we will get four files in the QED directory.
\begin{itemize}
\item[1.] lgrng1.mdl : This file contains the Feynman rules.
\item[2.] prtcls1.mdl: This file contains the list of particles and their properties, for example, spin, mass and width.
\item[3.] vars1.mdl: This file contains parameters that we defined, for example, the gauge coupling ``ee" and the 
mass of the electron ``me".
\item[4.] func1.mdl: If we want to introduce a function which depends on the above parameters, 
then this file contains the definition of this function.
\end{itemize}
Since these files are specific to {\sc CompHEP} and {\sc CalcHEP}, it would be good to consult their 
manual \cite{Belyaev:2012qa} and the TASI 2012 lecture note \cite{Kong:2012vg} for more detailed information.

\subsection{MC4BSM 2012 on-site exercise}

In the MC4BSM 2012 workshop tutorial, we work with the simple model described in Section~\ref{sec:model}.
In this section, we will show the additional Lagrangian terms for this model,
\begin{eqnarray}
&&\mathcal{L}_{\textrm{s.m}}= -\frac{m_1^2}{2} \phi_1^2- \frac{m_2^2}{2} \phi_2^2 -m_{12}^2 \phi_1 \phi_2 ,\\
&&\mathcal{L}_{\textrm{f.m}} = M_U \bar U U +M_E \bar{E} E ,\\
&&\mathcal{L}_{\textrm{Yuk}}= \lambda_1 \phi_1 \bar{U} P_R u + \lambda_2 \phi_2 \bar{U} P_R u 
				+ \lambda_1' \phi_1 \bar E P_R e+ \lambda_2' \phi_2 \bar E P_R e +h.c. ,
\end{eqnarray}
with corresponding kinetic terms. 
{\sc LanHEP} already contains a model file for the SM in the ``mdl" directory, 
thus it is convenient to start from this pre-existing model file, in our case, ``newsm.mdl" file.
To write down a model file using {\sc LanHEP}, we need to understand
\begin{itemize}
\item[1.] How to read a pre-defined model file and how to use this in our model file.
\item[2.] How to relate mass eigenstates to interaction eigenstates.
\item[3.] How to check whether our model file is written up correctly or not.
\end{itemize}

\subsubsection{Setup}

Let's name our file ``mc4bsm.mdl"  and save it in the ``mdl" directory since this directory contains files that we will be using.

\begin{lstlisting}[language=sh,frame=single]
read newsm.
keys CKMdim=1.
keys SMmassless=1.
keys gauge_fixing=Feynman.
 
%%%%%%%%%%%%%%%%%%%%%%%%%%%%%%%%%%%%%%%%
% Remove the 1st and 2nd generation mass
do_if SMmassless==1.
parameter Mm = 0 , Ms =0, Mc=0 .
end_if. 
%%%%%%%%%%%%%%%%%%%%%%%%%%%%%%%%%%%%%%%%

%%%%%%%%%%%%%%%%%%%%%%%%%%%%%%%%%%%%%%%%
% Make CKM a diagonal Matrix.
do_if CKMdim==1.
parameter  Vub=0, Vcb=0, Vtd=0, Vts=0, Vtb=1, Vud=1, Vus=0, Vcs=1, Vcd=0.
end_if.
%%%%%%%%%%%%%%%%%%%%%%%%%%%%%%%%%%%%%%%%

do_if gauge_fixing==Feynman.
    model 'MC4BSM_feynman'/4.
do_else_if gauge_fixing==unitary.
    model 'MC4BSM_unitary'/3. 
end_if. 
\end{lstlisting}

On the first line in this code, we see the ``read" command. This command allows us
to use a model file which is already defined in the ``mdl" directory, in our case ``newsm.mdl". 
There are also lines starting with ``keys" and below them a section with ``do\_if, end\_if". 
By using this method, we can control different set-ups for our model file.
If you look into ``newsm.mdl" file, the CKM entries, Gauge choice and masses of SM particles are already defined. 
But after we read this model file, we can over-write/replace those pre-defined parameters with new values of our choice.
We put the {\tt CKMdim} and {\tt SMmassless} flags since the {\sc FeynRules} 
example from Section \ref{sec:FR} uses restrictions on the CKM and Particle masses.
In above code, we can also see that ``\%"  is a flag to comment out a line.

\subsubsection{Particles and corresponding parameters}

We will follow the same conventions as the {\sc FeynRules} tutorial in Section~\ref{sec:FR}
(see Table~\ref{tab:particlesLH}). 

\begin{table}[ht]
\begin{center}
\begin{tabular}{|c||c|c|c|c|c|c|}
\hline
Fields &$U$ & $E$ & $\phi_1$& $\phi_2$ & $\Phi_1$& $\Phi_2$ \\
\hline
{\sc LanHEP} &\verb+uv+ & \verb+ev+ & \verb+pi1+ & \verb+pi2+ & \verb+p1+ &  \verb+p2+\\
\hline
\end{tabular}
\caption{\label{tab:particlesLH}Symbols used for the fields in the {\sc LanHEP} implementation.}
\end{center}
\end{table}

For the scalar fields,  
$(\Phi_1, \Phi_2)$ are the mass eigenstates (physical states) of  $(\phi_1, \phi_2)$ with mixing angle $\theta$:
\begin{equation}
\left(\begin{array}{c}
\phi_1 \\
\phi_2
\end{array}\right) = \left(\begin{array}{cc}
-\sin\theta & \cos\theta \\
\cos\theta & \sin\theta
\end{array}\right)\,\left(\begin{array}{c}
\Phi_1 \\
\Phi_2
\end{array}\right)\, ,
\end{equation}
coming from the eigen states of the following mass matrix, $\textrm{M}^2$,
\begin{equation}
\textrm{M}^2 = 
\left(\begin{array}{cc}
m_1^2 & m_{12}^2 \\
m_{12}^2 & m_2^2
\end{array}\right)\,.
\end{equation}

Thus there are five mass parameters, including the two eigenvalues (MPe1, MPe2) of the mass matrix, 
and a corresponding  mixing angle $\theta$, summarized in Table~\ref{tab:masses}.
\begin{table}[!h]
\begin{center}
\begin{tabular}{|c||c|c|c|c|c|c|}
\hline
Mass parameters &$m_1$ & $m_2$ & $m_{12}$& $M_U$ & $M_E$& $\tan(\theta)$ \\
\hline
{\sc LanHEP} &\verb+MM1+ & \verb+MM2+ & \verb+MM12+ & \verb+Muv+ & \verb+Mev+ &  \verb+tth+\\
\hline
\end{tabular}
\caption{\label{tab:masses}Symbols used for the mass parameters in the {\sc LanHEP} implementation.}
\end{center}
\end{table}

For coupling constants, we use the notation in Table~\ref{tab:parameters}.

\begin{table}[ht]
\begin{center}
\begin{tabular}{|c||c|c|c|c|}
\hline
Coupling constants &$\lambda_1$ & $\lambda_2$ & $\lambda_{1}'$& $\lambda_2'$   \\
\hline
{\sc LanHEP} &\verb+lam1+ & \verb+lam2+ & \verb+lam1p+ & \verb+lam2p+  \\
\hline
\end{tabular}
\caption{\label{tab:parameters}Symbols used for coupling parameters in the {\sc LanHEP} implementation.}
\end{center}
\end{table}

We sum up the above set-up for the particles in the following:
\begin{lstlisting}[language=sh,frame=single]
% Define new parameters.
parameter  MM1 = 200     : 'Mass parameter for pi1',
    MM2 = 300     : 'Mass parameter for pi2',
    MM12 = 50     : 'Mixing term for pi1 and pi2',
    lam1 = 1      : 'Yukawa interaction for pi1, U and u',
    lam2 = 1      : 'Yukawa interaction for pi2, U and u',
    lam1p = 1     : 'Yukawa interaction for pi1, E and e',
    lam2p = 1     : 'Yukawa interaction for pi2, E and e',
    Muv = 500.0   : 'Mass parameter for uv and v',
    Mev = 250.0   : 'Mass parameter for ev and e'.

parameter  
    MPe1 = sqrt(MM2**2+MM1**2-sqrt(4.0*MM12**4+(MM2**2-MM1**2)**2))/Sqrt2, 
    MPe2 = sqrt(MM2**2+MM1**2+sqrt(4.0*MM12**4+(MM2**2-MM1**2)**2))/Sqrt2.

parameter 
    tth = (MM2**2-MM1**2+sqrt(4.0*MM12**4+(MM2**2-MM1**2)**2))/(2.0*MM12**2).
parameter sth = tth/sqrt(1+tth**2), cth = 1.0/sqrt(1+tth**2).
angle sin=sth,cos=cth,tan=tth, texname='\\theta'.

% Define new particles.
spinor ev:('heavy electron',mass Mev, width Wev=auto),
       uv:('heavy quark', color c3, mass Muv, width  Wuv=auto).
     
scalar  p1/p1:('LNP',mass MPe1),
        p2/p2:('Heavy scalar',mass MPe2, width Wpe2=auto).
        
let pi1= -sth*p1+cth*p2, pi2=cth*p1+sth*p2.

prtcproperty pdg:(  ev=9000009,uv=9000008,p1=9000006,p2=9000007).
\end{lstlisting}

As we notice here, the mathematical commands are similar with Fortran, for example $x^2 \rightarrow  x**2$. 
For more information, we refer the user to the manual v.3.0 \cite{Semenov:2008jy}.
There are three major points that need to be explained: 
\begin{description}
\item[1.] Particle width: {\sc CompHEP}/{\sc CalcHEP} can calculate the width of a particle on the fly. 
Thus in the particle definition, we can set it to ``auto" and then {\sc CompHEP}/{\sc CalcHEP} will calculate it. 
\begin{itemize}
\item[] width Wev=auto
\end{itemize}
\item[2.] Physical state: We defined the physical states $(\Phi_1, \Phi_2)$ in the particle definitions
and can use the ``let" command to define interaction eigenstates: 
\begin{itemize}
\item[] let pi1= -sth*p1+cth*p2, pi2=cth*p1+sth*p2.  
\end{itemize}
We will define our Lagrangian in terms of interaction eigenstates and {\sc LanHEP} will automatically 
replace terms with the mass eigenstates. 
A Width of p1 is not defined, because p1 is a stable particle in our model. By default, its width $= 0$.
\item[3.] PDG number: We set the PDG number to be consistent with the model from {\sc FeynRules}. 
In this onsite exercise, we will not change the ``calchep.rc" file, but use the following command instead
\begin{itemize}
\item[] prtcproperty pdg:(  ev=9000009,uv=9000008,p1=9000006,p2=9000007).
\end{itemize}
\end{description}

\subsubsection{Setting up the Lagrangian}

We start this subsection with the code for the Lagrangian, which we will then explain step by step.
\begin{lstlisting}[language=sh,frame=single]
% Kinematics terms for new  fermion particles
% New fermion field is a counter part of SM right-handed particle 

let PR = (1+g5)/2.

lterm   anti(psi)*gamma*PR*(i*deriv - Y*g1*B1)*psi
 where 
 psi=ev, Y= -1;  
 psi=uv, Y=  2/3.

lterm  GG*anti(uv)*lambda*gamma*G*uv.

% Kinematic terms for scalars
lterm deriv*pi1*deriv*pi1/2+deriv*pi2*deriv*pi2/2.

% New mass term
lterm -MM1**2/2 * pi1*pi1- MM2**2/2 * pi2*pi2- MM12**2*pi1*pi2.
lterm -Muv*anti(uv)*uv-Mev*anti(ev)*ev.

% New Yukawa type interaction.
lterm lam1*pi1*anti(uv)*PR*u+lam2*pi2*anti(uv)*PR*u
      +lam1p*pi1*anti(ev)*PR*e+lam2p*pi2*anti(ev)*PR*e + AddHermConj.

% Checking a Lagrangian
CheckHerm.
\end{lstlisting}

\paragraph{Fermions.}
\begin{description} 
\item[1.] We defined a right-handed projector PR with g5 (g5 is defined in newsm.mdl as g5=$\gamma_5$.) 
\item[2.]  We used ``where" grammar to reduce unnecessary repeats and to give a clear structure.
\begin{itemize}
\item psi is a generic notation for fermions here. B1 is $U(1)_Y$ gauge field defined in ``newsm.mdl" file.
\item anti($\psi$) denotes $\bar{\psi}$, similarly to the {\sc FeynRules} conventions.
\item  Y is the hypercharge of the new particles, and g1 is defined to be the $U(1)_Y$ coupling in ``newsm.mdl".
\item For uv(colored particle) there is an interaction with G (Gluon) through the strong coupling GG defined in ``newsm.mdl".
\end{itemize}
\end{description}
For example,
\begin{table}[!h]
\begin{center}
\begin{tabular}{c||c}
Lagrangian & $\qquad \qquad \bar{\psi} \gamma^\mu \left(\partial_\mu-\textrm{Y}_{\psi} \,  g_1 B_{1\mu}\right) \psi $\\[2mm]
{\sc LanHEP} &  {\tt anti(psi)*gamma*PR*(i*deriv - Y*g1*B1)*psi}
\end{tabular}
\end{center}
\end{table}

Notice that we can omit the Lorentz indexes when they are summed up. Alternatively, as in the ``qed.mdl" file, we can write them
down explicitly:
 
\begin{lstlisting}
 anti(psi)*gamma^mu*PR*(i*deriv^mu - Y*g1*B1^mu)*psi
\end{lstlisting} 

\paragraph{Scalar particles.}
Usually, it is easy to write down the Lagrangian in terms of interaction eigenstates, instead of 
mass eigenstates (or physical states). 
Thus 
when we define Lagrangians, we use interaction eigenstates. During compiling of {\sc LanHEP}, the 
interaction states will be replaced with physical states and we will not see unphysical states in the
{\sc CompHEP}/{\sc CalcHEP} model files.

To summarize, the procedure is the following:
\begin{itemize}
\item[1.] Define physical states.
\item[2.] Relate unphysical states to the corresponding physical states using the ``let" command.
\item[3.] Describe Lagrangian in terms of unphysical states.
\end{itemize}

The same technique is applied in ``newsm.mdl" for gauge fields,
thus we recommend the user to read the ``newsm.mdl" file at this stage.

\subsubsection{Finalizing the Lagrangian script}
We add
\begin{lstlisting}
 + AddHermConj
\end{lstlisting} 
to the Yukawa interaction to get a hermitian conjugate term. To check our script for the Lagrangian, we add the
\begin{lstlisting}
CheckHerm.
\end{lstlisting} 
command to make sure that our Lagrangian is hermitian.  For additional commands for cross-checks,
see section 6 of the {\sc LanHEP} manual 2.0 \cite{Semenov:2008jy}.

 We can get a model file for {\sc CompHEP}/{\sc CalcHEP} by compiling in the {\sc LanHEP} directory 
\begin{lstlisting}[language=sh,frame=single]
mkdir tutorial
./lhep  -OutDir tutorial mdl/mc4bsm.mdl -CalcHEP
\end{lstlisting} 
so that we do not mix up our newly created model files with existing ``mdl" files.


\section{{\sc MadGraph} tutorial}
\label{sec:MG}

\subsection{Preamble}

\madgraph5 \cite{Alwall:2011uj} can be run on a local computer or via the web at one of the following website:
\begin{itemize}
\item http://madgraph.hep.uiuc.edu
\item http://madgraph.phys.ucl.ac.be
\end{itemize}
The registration is straightforward, and you can instantly creates optimized code for the computation of the cross-section of
any processes. That code can be  run directly on the web or downloaded and run locally. However, for security reason, generating events 
on the web is allowed only after that you have sent an email to one of the authors of \madgraph5.
Since most of the functions are available on the internet,  most users will not need to install \madgraph5.

The \madgraph5 collaboration plans to continue to improve this code both by adding new functionalities and by making the code  easier to use.
 In particular the way to compute decay widths presented in this paper will soon be improved.
 New tutorials and updates on this tutorial can be found at \cite{MGupdates}.

\subsection{Installation of \madgraph5}

\subsubsection{Installation on Windows}
 
\madgraph5 and the  associated programs are designed and tested on Linux and MacOs operating systems. The windows compatibility via \texttt{cygwin} is currently not supported. 
For Windows user, we advise they install Linux in dual boot. Virtual machines are another possibility. Note that some virtualbox packages do not include the library readline.
This library is not mandatory  but enables the auto-completion (See the paragraph associated to python installation to learn how to solve this).

\subsubsection{Installation on Linux / MacOS}
\paragraph{MadGraph 5}
    The last version of \madgraph5 can be found at the following page: https://launchpad.net/madgraph5.
    This website is also the place, where you can ask question, make suggestions or report a bug.
    The installation is straightforward since you have only to untar it
    \begin{verbatim}
      tar -xzpvf MadGraph5_v1.X.Y.tgz
    \end{verbatim}
    No compilation are required for \madgraph5, you can just launch it.
    \begin{verbatim}
    ./MadGraph5_v1.X.Y/bin/mg5
    \end{verbatim}
    If you don't have a valid python version, \madgraph~5 will crash directly with an 
    explicit message. In this case, you will need to install \python2.7.    
    
    If you have admin rights on your system, you can run the following command:
    \begin{verbatim}
     sudo ln -s MadGraph5_v1.X.Y/bin/mg5 /usr/local/bin
    \end{verbatim} 
    such that \madgraph5 can be launched from any directory.
       
\paragraph{Python}
    The only requirement for \madgraph5 is to have a current version of  \python~(version 2.6 or 2.7). 
    In most cases, it can be installed via your favorite repository manager.
    However, some of the linux repository distributes python 2.5. In that case, you will need to
    download python at the following link:
    http://www.python.org/download/
    and follows the associate instructions.
    
    Note that for some linux versions (especially in virtual machine), the library readline is not present on the system. 
    This will disable the auto-completion. If you care about that point, you will should first install that library (via your repository)
    and then to recompile python from the source code.
    
    \paragraph{Optional package}
    Various optional packages can be linked to \madgraph5 in order to customize the output, create plots, ...
    The installation of those packages is easy since you can install them by launching mg5 and typing
       \begin{verbatim}
     mg5> install NAME
    \end{verbatim}  
    Where {\sc NAME} is one of the following package name:
    \begin{itemize}
    \item MadAnalysis: A package to draw automatically various histogram linked to the event generation.
    \item ExRootAnalysis \cite{Alwall:2007st}: A package to convert the various output in a {\sc ROOT} format.
    \item pythia-pgs: A package containing \pythia6 \cite{Sjostrand:2006za} and \pgs~\cite{PGS4}. \pythia6  is able to shower and to hadronize your events and is able to perform the matching \cite{Alwall:2008pm, Sjostrand:2006za} for multi-jet production. \pgs~is a fast detector simulation package.
    \item Delphes \cite{Ovyn:2009tx}: A package allowing to have a fast detector simulation, in replacement of \pgs.
    \end{itemize}
    Note that some of those programs might have some extra-dependencies (especially in Root).
    
\paragraph{Additional instructions for MacOs}

Compared to Linux the installation on MacOS might be more complex since  MacOS doesn't provide various set of
default programs present on Linux. Two important programs which are not present by default  are gmake and gfortran4.x.
We advise you to first check if those program are install via the commands.
\begin{verbatim}
      $> make --version
      $> gfortran --version
\end{verbatim} 
In order to install gmake it is easiest to install xcode (free but requiring an apple developer account) 
\begin{itemize}
\item MacOs 10.5:  https://connect.apple.com/cgi-bin/WebObjects/\\
MemberSite.woa/wa/getSoftware?bundleID=20414
\item MacOs 10.6:   http://connect.apple.com/cgi-bin/WebObjects/\\
                  MemberSite.woa/wa/getSoftware?bundleID=20792
\item MacOs 10.7: http://itunes.apple.com/us/app/xcode/id448457090?mt=12
\end{itemize}
Concerning gfortran you can download it from the following link:
\begin{itemize}
\item MacOs 10.5:   http://prdownloads.sourceforge.net/hpc/ \\
gcc-lion.tar.gz?download 
\item MacOs 10.6:   http://prdownloads.sourceforge.net/hpc/\\
gcc-snwleo-intel-bin.tar.gz?download
\item MacOs 10.7:  http://sourceforge.net/projects/hpc/files/hpc/gcc/\\
gcc-leopard-intel-bin.tar.gz/download
\end{itemize}

\subsubsection{Testing the installation and learning the syntax}

\madgraph5 includes a build-in tutorial.
\begin{verbatim}
$> ./bin/mg5
mg5> tutorial
\end{verbatim}
Then just follow the instructions on the screen and you will learn the basic command/usage of \madgraph5. This takes around 15-20 minutes.

\subsection{MC4BSM exercise}

In this section, we will  study a full example on how to generate events for BSM theories.
We will assume that you have your own UFO model \cite{Degrande:2011ua} according to the {\sc FeynRules} \cite{Christensen:2008py} tutorial.
If you don't, you can download the associated model at the following address:\\
http://feynrules.irmp.ucl.ac.be/attachment/wiki/WikiStart/MC4BSM\_2012\_UFO.tgz
We will split this example into three sections. First we will show how you can test the validity of the model,
second, we will present how to create a valid set of parameters associated with the model,
and finally we will present how you can generate BSM events, both with and without the associated decays.

\subsubsection{Importing and checking the model}

     The simplest way to have access to a model in MG5 is to put it in the directory: MG5\_DIR/models
     after that you can simply import it by typing 
\begin{verbatim}
    mg5> import model MC4BSM_2012_UFO
\end{verbatim} 
     or 
\begin{verbatim}
   mg5> import model MC4BSM_2012_UFO --modelname
\end{verbatim} 
The option --modelname tells MG5 to use the name of the particles defines in the UFO model, and not the usual MG5 conventions for the particles of the SM/MSSM.
For this particular model, this changes only the name associated to $\tau$ lepton (ta- and tt- respectively).     

If you have developed your own model following the {\sc FeynRules} tutorial from Section~\ref{sec:FR}, this will be the first time that you are going to use this model. It is therefore crucial to start by checking the model.
\madgraph5 performs some sanity checks the first time that you load the model, but those test are quite weak. We therefore suggest to test,  three properties for on a series of processes. 
\begin{itemize}
\item The gauge invariance, by testing the Ward identities.
\item The Lorentz invariance.
\item The \aloha~\cite{deAquino:2011ub} consistency, by evaluating the same square matrix element by different set of Helicity amplitudes.
\end{itemize}
For instance, we present how to check those properties for all the $2\to2$ BSM particles productions:
\begin{verbatim}
   mg5> import model MC4BSM_2012_UFO
   mg5> define new = uv uv~ ev ev~ p1 p2
   mg5> check p p > new new
\end{verbatim}
which results in the following output:
\begin{verbatim}
Gauge results:
Process          matrix           BRS              ratio            Result
g g > uv uv~     7.4113020914e-01 1.0055761722e-31 1.3568144434e-31 Passed
g u > uv p1      3.8373877873e-02 6.1629758220e-33 1.6060341471e-31 Passed
g u > uv p2      2.5726129500e-02 4.4176419953e-33 1.7171809678e-31 Passed
g u~ > uv~ p1    2.0117011717e-01 8.3456964257e-34 4.1485766093e-33 Passed
g u~ > uv~ p2    1.9705216573e-01 1.8809915790e-32 9.5456529090e-32 Passed
Summary: 5/5 passed, 0/5 failed
Lorentz invariance results:
Process          Min element      Max element      Relative diff.   Result
g g > uv uv~     4.8743514998e-01 4.8743514998e-01 1.1388417769e-16 Passed
g u > uv p1      1.3897148577e-01 1.3897148577e-01 1.1983282238e-15 Passed
g u > uv p2      5.7988729593e-02 5.7988729593e-02 5.9829676840e-16 Passed
g u~ > uv~ p1    8.8655875905e-03 8.8655875905e-03 1.7610238605e-15 Passed
g u~ > uv~ p2    1.0373598029e-01 1.0373598029e-01 8.0267932700e-16 Passed
u u > uv uv      2.1536620557e+00 2.1536620557e+00 1.8558171084e-15 Passed
u u~ > uv uv~    8.4033626775e-01 8.4033626775e-01 2.1138643036e-15 Passed
u u~ > p1 p1     7.6001259005e-03 7.6001259005e-03 2.9672410014e-15 Passed
u u~ > p1 p2     9.5102790993e-05 9.5102790993e-05 2.7645774018e-14 Passed
u u~ > p2 p2     4.6427735529e-04 4.6427735529e-04 8.6404128750e-15 Passed
u u~ > ev ev~    2.3919237366e-03 2.3919237366e-03 2.7196573767e-15 Passed
c c~ > uv uv~    8.5752329113e-01 8.5752329113e-01 2.3304340127e-15 Passed
d d~ > uv uv~    8.9275046680e-01 8.9275046680e-01 3.2333557600e-15 Passed
d d~ > ev ev~    4.8145327614e-04 4.8145327614e-04 5.6298410782e-16 Passed
s s~ > uv uv~    8.8941849408e-01 8.8941849408e-01 3.6199458330e-15 Passed
u~ u~ > uv~ uv~  2.3800428911e+00 2.3800428911e+00 1.8658874237e-15 Passed
Summary: 16/16 passed, 0/16 failed
Not checked processes: c c~ > ev ev~, s s~ > ev ev~
Process permutation results:
Process          Min element      Max element      Relative diff.   Result
g g > uv uv~     4.9184278197e-01 4.9184278197e-01 2.2572721717e-16 Passed
g u > uv p1      4.1859552985e-02 4.1859552985e-02 3.3153215498e-16 Passed
g u > uv p2      2.0129184319e-01 2.0129184319e-01 1.1030978772e-15 Passed
g u~ > uv~ p1    9.5566536137e-02 9.5566536137e-02 5.8086390357e-16 Passed
g u~ > uv~ p2    3.6165811126e-03 3.6165811126e-03 3.8372671248e-15 Passed
u u > uv uv      2.1101603787e+00 2.1101603787e+00 2.1045282356e-15 Passed
u u~ > uv uv~    1.3549964258e+00 1.3549964258e+00 1.1470969258e-15 Passed
u u~ > p1 p1     4.9555140623e-03 4.9555140623e-03 1.4002369515e-15 Passed
u u~ > p1 p2     2.0863569608e-02 2.0863569608e-02 6.6516842845e-16 Passed
u u~ > p2 p2     1.2914424155e-03 1.2914424155e-03 4.7013572521e-15 Passed
u u~ > ev ev~    1.7823584674e-03 1.7823584674e-03 0.0000000000e+00 Passed
c c~ > uv uv~    9.2608997797e-01 9.2608997797e-01 1.0789456164e-15 Passed
d d~ > uv uv~    8.3532448258e-01 8.3532448258e-01 1.3290919251e-16 Passed
d d~ > ev ev~    5.8126525280e-04 5.8126525280e-04 9.3262255680e-16 Passed
u~ u~ > uv~ uv~  2.1759606129e+00 2.1759606129e+00 2.0408880897e-16 Passed
Summary: 15/15 passed, 0/15 failed
\end{verbatim}

More informations about these checks (like the values of the random phase-space points) can be obtained via the commands:
\begin{verbatim}
    mg5> display checks
\end{verbatim}
The display commands is very useful in order to obtained information on the particles, couplings, interactions, ...
For example\footnote{Note that the spin is written in the $2S+1$ convention.}

\begin{verbatim}
     mg5>display particles 
        Current model contains 21 particles:
        ve/ve~ vm/vm~ vt/vt~ e-/e+ m-/m+ tt-/tt+ u/u~ c/c~ t/t~ d/d~ s/s~ 
        b/b~ w+/w- uv/uv~ ev/ev~ a z g h p1 p2
     mg5>display particles p1
       Particle p1 has the following properties:
       {
          'name': 'p1',
          'antiname': 'p1',
          'spin': 1,
          'color': 1,
          'charge': 0.00,
          'mass': 'MPe1',
          'width': 'Wpe1',
          'pdg_code': 9000006,
          'texname': 'p1',
          'antitexname': 'p1',
          'line': 'dashed',
          'propagating': True,
          'is_part': True,
          'self_antipart': True
       }
     mg5>
\end{verbatim}

\subsubsection{Computation of the width and of the branching ratio}

     The first requirement in order to have valid event generation is to have a valid set of input parameters. This includes having the correct widths and branching ratios (needed by {\sc Pythia}). Currently, \madgraph5 is not able to compute these automatically and requires that you prescribe {\bf all}  decay channels. Note that this limitation will be resolve in a near future.

So with the current version of \madgraph, you need to do the following in order to compute the width:
\begin{verbatim}
	mg5> import model MC4BSM_2012_UFO
	mg5> generate uv > u p1
	mg5> add process uv > u p2
	mg5> add process p2 > ev e+
	mg5> add process p2 > ev~ e-
	mg5> add process ev > e- p1
	mg5> output
	mg5> launch
\end{verbatim}
The last command will ask you if you want to change the mass spectrum and/or the cuts. 
In principle all cuts should be set to zero for width computation.
Then your browser will automatically open an html page with the results.\footnote{You can de-activated this behavior by modifying the configuration files
present at the following path: ./input/mg5\_configuration.txt}  One of the results will be the new param\_card.dat.
Note that this script modified {\bf only} the width of the particles present in the initial states of the generated processes. 
This means that the width of $p_1$ {\bf is not modified} and therefore is still has it default value ($1.0$) which is not correct.
You have to modify this card manually before using it for an event generation.  

\subsubsection{Generating events}

In this section, we start the computation of the cross-sections and the generation of events for the proposed process of interest:
$$
 p p \to U \bar U
$$
First we will generate this exact process, \pythia~being in charge of the decays.
Note that in this way, you lose the full spin-correlations.
\begin{verbatim}
   import model MODELNAME
   generate p p  > uv uv~
   output
   launch
\end{verbatim}
If you have install the {\sc pythia-pgs}/\delphes~package, you will be asked if you want to run the package.
Next a question will ask if you want to edit one of the cards. At this stage, you can enter the path of the card that you have created in the previous 
step and/or edit those manually.

In this case, the number of decay channels is quite limited, it's therefore possible to prescribe all the steps of the decay chain.
The syntax for the decay chains is the following:
the production first, then the decays of the final states separate by a comma. To avoid ambiguity parenthesis should
be present in the case of sub-decay.
\begin{verbatim}
import model MC4BSM_2012_UFO
generate p p > uv uv~ , uv > u p1, uv~ > u~ p1
define l = e+ e-
define lv = ev ev~
add process p p > uv uv~, uv > u p1, \
                         (uv~ > u~ p2, (p2 > l lv, lv > l p2))
add process p p > uv uv~, uv~ > u~ p1,\
                         (uv > u p2, (p2 > l lv, lv > l p2))
add process p p > uv uv~, (uv > u p2, (p2 > l lv, lv > l p2)), \
                          (uv~ > u~ p2, (p2 > l lv, lv > l p2))
output
launch
\end{verbatim}

The validity of this calculation must be considered, since specifying the decay sequence means the contribution of non-resonance Feynman diagram have been neglected.
This is however valid since the interferences with those diagrams are negligible if the intermediate particles are on-shell. 
In order to ensure such a condition, we have associate to each of the decaying particles an additional cuts called \emph{BW\_cut}:
$$
| m_{virt} - m_0| < BW_{cut} * \Gamma
$$
This cut can be specified in the run\_card.dat. 

If you have installed the {\sc MadAnalysis} package, you should have automatically distributions created at parton level, after \pythia~and at the reconstructed level --if runned--. 
Those one can be used to check the sanity of the productions but also to search the potential observables of this production.

\section{{\sc CalcHEP} tutorial}
\label{sec:CH}

\subsection{Installation of {\sc CalcHEP}}

Download {\sc CalcHEP} from:
\begin{center}
\begin{verbatim}
http://theory.sinp.msu.ru/~pukhov/CALCHEP/calchep_3.4.0.tgz
\end{verbatim}
\end{center}

Move it to the directory where you would like to do the tutorial.  For example\footnote{If your OS already unzipped the file, change the endings appropriately.}:
\begin{verbatim}
mv ~/Downloads/calchep_3.4.0.tgz \
~/physics/projects/MC4BSM-2012/Sandbox/calchep_3.4.0.tgz
\end{verbatim}

Change to that directory.  For example:
\begin{verbatim}
cd ~/physics/projects/MC4BSM-2012/Sandbox
\end{verbatim}

Unpack the {\sc CalcHEP} code\footnote{If your OS already unzipped, then you would do tar xvf calchep\_3.4.0.tar  Or, if your OS already fully unpacked the file, you can skip this step.}:
\begin{verbatim}
tar xvzf calchep_3.4.0.tgz
\end{verbatim}

Change to the codes directory:
\begin{verbatim}
cd calchep_3.4.0
\end{verbatim}

Compile the code\footnote{Note that you need the X11 development libraries installed on your system to use the graphical user interface (gui).  For Mac, these used to come on the X11 disk.  Now, I assume you get it through the app store.  For Linux, your distribution likely has it in a repository.  Google X11 development libraries and your distribution to see how to install it.  However, note that it is not strictly necessary to have the gui working for this tutorial.}:
\begin{verbatim}
make
\end{verbatim}

Create a working directory for your calculations\footnote{mkUsrDir is planned to change to mkWorkDir.}:
\begin{verbatim}
./mkUsrDir ../ch_3.4.0
\end{verbatim}

Change directory to the working directory:
\begin{verbatim}
cd ../ch_3.4.0
\end{verbatim}

If you have the X11 development libraries installed, you can start the symbolic session gui (otherwise skip this step):
\begin{verbatim}
./calchep &
\end{verbatim}

Feel free to try some processes out and play.  This will not be part of the tutorial.

\subsection{Install the MC4BSM 2012 model}
Download the {\sc CalcHEP} model files from:
\begin{center}
\begin{verbatim}
http://feynrules.irmp.ucl.ac.be/raw-attachment/wiki/WikiStart/MC4BSM_2012-CH.tgz
\end{verbatim}
\end{center}

Move this file to a convenient location.  For example\footnote{Again, change the ending as appropriate.}:
\begin{verbatim}
mv ~/Downloads/MC4BSM_2012-CH.tgz \
~/physics/projects/MC4BSM-2012/Sandbox/MC4BSM_2012-CH.tgz
\end{verbatim}

Change to this directory:
\begin{verbatim}
cd ~/physics/projects/MC4BSM-2012/Sandbox
\end{verbatim}

Unpack model files (if not already done):
\begin{verbatim}
tar xzvf MC4BSM_2012-CH.tgz
\end{verbatim}

Install the model\footnote{The model can be installed more easily using the gui, but I will describe the process that will work for everyone.}.  We need to change the model number to one that is not already used.  Since we just freshly installed {\sc CalcHEP}, we know there are only three models included by default (the Standard Model with general CKM matrix, the Standard Model with diagonal CKM matrix and the Standard Model with effective Higgs vertices).  The model number is given by the integer in the model file names.  We will change it to 4.  

Change to the model file directory:
\begin{verbatim}
cd MC4BSM_2012-CH
\end{verbatim}

Change the file names in such a way that the 1 becomes a 4.
\begin{verbatim}
mv prtcls1.mdl prtcls4.mdl
mv vars1.mdl vars4.mdl 
mv func1.mdl func4.mdl 
mv lgrng1.mdl lgrng4.mdl 
mv extlib1.mdl extlib4.mdl 
\end{verbatim}

Now, copy the model files to the models subdirectory of the {\sc CalcHEP} work directory:
\begin{verbatim}
cp *4.mdl ../ch_3.4.0/models/.
\end{verbatim}

Change directory back to the {\sc CalcHEP} working directory:
\begin{verbatim}
cd ../ch_3.4.0
\end{verbatim}

If you have the X11 development libraries installed, you can start the symbolic session gui (otherwise skip this step):
\begin{verbatim}
./calchep &
\end{verbatim}
You should see the tutorial model as the third model in the list.  You can check your model for syntactical correctness, change it, calculate cross sections and generate distributions.  Feel free to play.  This will not be part of the tutorial.

\subsection{Generate events}
We, next, write a batch file to generate events.

Open your favorite text editor (emacs, vi, gedit,...).  For example:
\begin{verbatim}
emacs &
\end{verbatim}

At the top of the file, we will choose the model.  Add the lines:
\begin{verbatim}
Model:         MC4BSM_2012
Gauge:         Feynman
\end{verbatim}
The model name can be found at the top of the file \verb|models/prtcls3.mdl|.

We next specify the process we are interested in:
\begin{verbatim}
Process:    p,p->uv,uv~
Decay:      uv->u,p1
Decay:      uv~->u~,p1
Decay:      uv->u,p2
Decay:      uv~->u~,p2
Decay:      p2->e-,ev~
Decay:      p2->e+,ev
Decay:      ev->e-,p1
Decay:      ev~->e+,p1
Alias:      p=u,u~,d,d~,G
\end{verbatim}

We next specify the parton distribution function.  We will use the built in CTEQ 6L set:
\begin{verbatim}
pdf1:      cteq6l (proton)
pdf2:      cteq6l (proton)
\end{verbatim}

Specify the momenta:
\begin{verbatim}
p1:        4000
p2:        4000
\end{verbatim}

Specify the number of events to generate and the filename to use:
\begin{verbatim}
Number of events (per run step): 10000
Filename:                        MC4BSM
\end{verbatim}

Specify your parallelization info:
\begin{verbatim}
Parallelization method:      local
Max number of cpus:          2
sleep time:                  1
\end{verbatim}
I entered 2 for the max number of cpus because I have a dual core machine.  You should change this as appropriate.

Specify how many times Vegas should run:
\begin{verbatim}
nSess_1:   5
nCalls_1:  10000
nSess_2:   5
nCalls_2:  100000
\end{verbatim}
The first five runs improve the grid.  Then the statistics are cleared, the grid is frozen and 5 more runs are done to calculate the cross section, widths and prepare the event generator.

After you have all of these pieces entered, save the file.  Call it  ``batch\_file" in your work directory.

Now, run the batch program by:
\begin{verbatim}
./calchep_batch batch_file
\end{verbatim}

The batch program will begin by stating where you can find information about the progress of your job.  For example, 
\begin{verbatim}
Processing batch:
Progress information can be found in the html directory.
Simply open the following link in your browser:
file:///Users/neil/physics/projects/MC4BSM-2012/Sandbox/ch_3.4.0/html/index.html
You can also view textual progress reports in /Users/neil/physics/projects/MC4BSM-2012/Sandbox/ch_3.4.0/html/index.txt
	and the other .txt files in the html directory.
Events will be stored in the Events directory.
\end{verbatim}

Note, the line that begins \verb|file:|.  You can copy and paste this line into your web browser's url bar.  This shows the progress of your calculation.  There is also a link on the left called \verb|Help| which contains detailed instructions about how to write batch files.  There are many options that we did not discuss in this tutorial.

When the program finishes, you will have the following file:
\begin{verbatim}
Events/MC4BSM-single.lhe.gz
\end{verbatim}
which contains your events in LHE format.

Enjoy!

\section{{\sc Pythia~8} tutorial}
\label{sec:PY}

\subsection{Introduction}

This exercise corresponds to the {\sc Pythia}~8 part of the 
more general MC tutorial given at the MC4BSM workshop at Cornell 
University 2012 \cite{MC4BSM}. It is for this reason focused 
on the particular task within this tutorial, however, should still 
serve as a good starting point to get familiar with the basics of 
how to use the \textsc{Pythia}~8 event generator. Much of these 
instructions were based on earlier \textsc{Pythia}~8 tutorials, 
which can be found on the homepage \cite{pythiaweb} and often 
include additional material than what is covered here.

Within this first exercise it is not possible to describe the 
physics models used in the program; for this we refer to the 
online manual \cite{pythiaman}, \textsc{Pythia}~8.1 brief 
introduction \cite{Sjostrand:2007gs}, to the full \textsc{Pythia}~6.4 
physics description \cite{Sjostrand:2006za}, and to all the further 
references found in them.

Finally, a good way to continue after the tutorial is often to 
chose a particular physics study your interest and then start 
to explore the different simulation and analysis aspects, using 
the different example programs together with the online manual, 
along the lines of the tutorial.

\subsection{Installation and pre-workshop exercise}
\label{sec:PYpreworkshop}

\textsc{Pythia}~8 is, by today's standards, a small package. It 
is completely self-contained, and is therefore easy to install 
for standalone usage, e.g. if you want to have it on your own 
laptop, or if you want to explore physics or debug code without 
any danger of destructive interference between different libraries. 

Denoting a generic \textsc{Pythia}~8 version \texttt{pythia81xx}
(at the time of writing \texttt{xx} = 62), here is how to install 
\textsc{Pythia}~8 on a Linux/Unix/MacOSX system as a standalone 
package. 

\begin{Enumerate}
\item In a browser, go to\\
\hspace*{10mm}\texttt{http://www.thep.lu.se/}$\sim$%
\texttt{torbjorn/Pythia.html}
\item Download the (current) program package\\
\hspace*{10mm}\texttt{pythia81xx.tgz}\\
to a directory of your choice (e.g. by right-clicking on the link).
\item In a terminal window, \texttt{cd} to where \texttt{pythia81xx.tgz} 
was downloaded, and type\\
\hspace*{10mm}\texttt{tar xvfz pythia81xx.tgz}\\
This will create a new (sub)directory \texttt{pythia81xx} where all
the \textsc{Pythia} source files are now ready and unpacked.
\item Move to this directory (\texttt{cd pythia81xx}) and do a
\texttt{make}. This will take $\sim$3 minutes
(computer-dependent). The \textsc{Pythia}~8 libraries are now
compiled and ready for physics. 
\item For test runs, \texttt{cd} to the \texttt{examples/} subdirectory. 
An \texttt{ls} reveals a list of programs, \texttt{mainNN}, with
\texttt{NN} from \texttt{01} through \texttt{28} (and beyond). These 
example programs each illustrate an aspect of \textsc{Pythia}~8. 
For a list of what they do, see the ``Sample Main Programs'' page 
in the online manual (point 6 below).\\ 
In order to test that the program installed successfully, most examples 
are suitable, however, for this tutorial (using version 8.162) 
a good choice would be \texttt{main11.cc}, which reads in a test LHEF 
called \texttt{ttbar.lhe}.\\
To execute this test program, do\\
\hspace*{10mm}\texttt{make main11}\\
\hspace*{10mm}\texttt{./main11.exe}\\
The output is now just written to the terminal, \texttt{stdout}, and 
if everything worked one should see (specific for main11.cc) a text 
based histogram, showing the a charge particle multiplicity, at the 
end of the program output.
To save the output to a file instead, do 
\texttt{./main11.exe >  main11.out}, after which you can study the 
test output at leisure by opening \texttt{main11.out}. See Appendix 
A for a brief explanation of the event record.
\item If you use a web browser to open the file\\
\hspace*{10mm}\texttt{pythia81xx/htmldoc/Welcome.html}\\
you will gain access to the online manual, where all available methods 
and parameters are described. Use the left-column index to navigate 
among the topics, which are then displayed in the larger right-hand 
field. 
\end{Enumerate}

\subsection{On-site exercise}

The task at hand for the \textsc{Pythia}~8 part of the tutorial 
is to read in the events, provided in LHEF format from the 
earlier steps, and continue the event simulation following the 
hard scatter generation. Given that the LHEF includes both the 
information for initialization and of the actual events, this 
is very straight forward. Copy first the LHEF 
\texttt{events\_with\_decay.lhe.gz} into the \texttt{examples} 
directory and unpack it using,\\ 
\hspace*{10mm}\texttt{gunzip events\_with\_decay.lhe.gz}\\
The LHEF should now be readable just like any ASCII file.

The BSM model as well as the hard process is described in detail 
in the earlier steps of this tutorial, but as a short reminder 
the events consists of, 
$pp \rightarrow U\bar{U}$, production at the LHC with $E_{cm} = 8$ 
TeV. The $U$ fermions have a mass of 500 GeV and the two following 
decay scenarios are possible:
\begin{eqnarray}
U \rightarrow u~\Phi_1; \\
U \rightarrow u~\Phi_2 \rightarrow u~e~E \rightarrow u~e~e~\Phi_1.
\end{eqnarray}

The second step is to create our own main program. Open a new file 
\texttt{mymain.cc} in the \texttt{examples} subdirectory with 
a text editor, e.g.\ Emacs. Then type the following lines (here 
with explanatory comments added):
\begin{verbatim}
     // Headers and Namespaces.
     #include "Pythia.h"      // Include Pythia headers.
     using namespace Pythia8; // Let Pythia8:: be implicit.

     int main() {             // Begin main program.

       // Set up generation.
       Pythia pythia;         // Declare Pythia object
       pythia.readString("Beams:frameType = 4"); // Beam info in LHEF.
       pythia.readString("Beams:LHEF = events_with_decay.lhe"); 
       pythia.init(); // Initialize; incoming pp beams is default.

       // Generate event(s).
       for (int iEvent = 0; iEvent < 1; ++iEvent) {
         if (!pythia.next()) {
           if (pythia.info.atEndOfFile()) break; // Exit at enf of LHEF.
           continue; // Skip event in case of problem.
         }
       } 

       pythia.stat();  // Print run statistics.
       return 0;
     }
\end{verbatim}

This will use one event from the LHEF and pass it through the remaining 
{\sc Pythia} 8 simulation. Go through the lines in the program and try to 
understand them by consulting the online manual \cite{pythiaman}. 

Next you need to edit the \texttt{Makefile} (the one in the 
\texttt{examples} subdirectory) so it knows what to do with 
\texttt{mymain.cc}. The lines\\
\hspace*{10mm}\texttt{\# Create an executable for one of the normal 
test programs}\\
\hspace*{10mm}\texttt{main00  main01 main02 main03 ... main09 main10 
main10}~\verb+\+ \\
and the three next enumerate the main programs that do not need 
any external libraries. Edit the last of these lines to include also 
\texttt{mymain}:\\
\hspace*{10mm}\texttt{        main31 ... main40 mymain:}~\verb+\+

Now it should work as before with the other examples:\\
\hspace*{10mm}\texttt{make mymain}\\
\hspace*{10mm}\texttt{./mymain.exe > mymain.out}\\
whereafter you can study \texttt{mymain.out}, especially the 
example of a complete event record (preceded by initialization 
information and by kinematical-variable listing for the same 
event). For this reason Appendix A contains a brief overview of 
the information stored in the event record. 

At this point you have in principle achieved your goal, the first 
event in the LHEF have been fully simulated by \textsc{Pythia}~8, 
using the default values of all settings. On the other hand, and 
despite only having one event, we only have access to the 
information of the individual particles in the event record, so 
interesting for technical validation but not much of real physics 
interest yet.

\subsection{A simple jet analysis}

We will now gradually expand the skeleton \texttt{mymain} program 
from above, in order to make a simple analysis, including jet 
reconstruction. The common standard, used in the experimental 
communities, is to produce fully simulated MC events in the 
\texttt{HepMc} format, either for analysis or further detector 
simulation. However, in order to keep the software infrastructure 
at a minimum we will in this part only use the analysis 
functionality already available within \textsc{Pythia}~8. For 
further information regarding how to produce \texttt{HepMc}, the 
reader is referred to the online manual \cite{pythiaman} together 
with other tutorials on the home page \cite{pythiaweb}. The final 
version of \texttt{mymain.cc}, with all steps included, is kept in 
Appendix B. Hence one can look there in case the instructions are 
unclear, however, that should of course be considered as the last 
resort, if getting completely stuck.

The BSM events generated in this tutorial will always include jets, 
from the $U$ decay, and missing $\Phi_1$ particles. They will some 
times also contain electrons, depending on the $U$ decay channel. 
For this reason we start with reconstructing jets from the final 
state particles in the event. For this we use the \textsc{Pythia}~8 
\texttt{SlowJet} program (found in the manual under \textit{Event 
Analysis}), which is a simpler version of the commonly used \texttt{FastJet} 
jet finder. The \texttt{SlowJet} program supports both the kT, anti-kT, 
and Cambridge/Aachen algorithms, where we will use the anti-kT which 
is a common choice within the LHC experiments. For further information 
about the jet reconstruction algorithms we recommend the online 
manual together with its references to the \texttt{FastJet} program.

Insert the following line before the event loop, in order to create a 
\texttt{SlowJet} object,\\ 
\hspace*{10mm}\texttt{SlowJet::SlowJet sJets(-1, 0.4, 50., 5., 1, 2);}\\
The first argument specifies the anti-kT algorithm, the second is the 
related R measure which roughly corresponds to the radius of the jet 
cone in ($y$,$\phi$). The next two corresponds to, minimum $p_T$ and 
maximum $|\eta|$, acceptance cuts usually implied by the experiment 
of interest. The last two arguments specifies respectively to use all 
final state particles and to use their actual mass in the reconstruction. 
The selection of which particles to include is normally done more 
carefully, however, in a trade--off between illustrating the physics 
without spending too much time on technical details and infrastructure, 
we chose a rather unsophisticated approach here. Just after this line, 
you should also create the following two histograms,\\
\hspace*{10mm}\texttt{Hist nJ( "Nr Jets", 10, -0.5, 9.5);}\\
\hspace*{10mm}\texttt{Hist j1pT( "Leading Jet pT", 100, 0., 1000.);}\\
which corresponds to our final analysis results.

The next step is to reconstruct jets from the final state particles in 
each event according to the jet algorithm we just defined. After that 
we also want to fill our histograms, once per event, and at the end of 
the program print them to our output. To do this, include the following 
lines inside the event loop, but at the end so that \texttt{pythia.next()} 
has been executed,
\begin{verbatim}
    // Jet analysis.
    if (sJets.analyze(pythia.event)) {
      nJ.fill(sJets.sizeJet());
      j1pT.fill(sJets.pT(0));
    }
\end{verbatim}
The \texttt{pythia} member \texttt{.event} corresponds to the event 
record and you can try to find out the meaning of these lines by using 
the online manual. To write the histograms to the output include the 
following line at the very end, just before the \texttt{return} 
statement,\\
\hspace*{10mm}\texttt{std::cout << nJ << j1pT;}\\

Now increase the number of events in the event loop, \textit{e.g.} to 
1000, compile and run the program. At the end of the output one should 
now see the number of jets and leading jet $p_T$ distributions, in 
traditional text based histograms. 

During the run you may receive problem messages. These come in three 
kinds:
\begin{Itemize}
\item a \textit{warning} is a minor problem that is automatically fixed 
by the program, at least approximately;
\item an \textit{error} is a bigger problem, that is normally still 
automatically fixed by the program, by backing up and trying again;
\item an \textit{abort} is such a major problem that the current
 event could not be completed; in such a rare case \texttt{pythia.next()} 
is \texttt{false} and the event should be skipped.  
\end{Itemize}
Thus the user need only be on the lookout for aborts. During event
generation, a problem message is printed only the first time it occurs. 
The above-mentioned \texttt{pythia.stat()} will then tell you how many
times each problem was encountered over the entire run.

What we did so far is clearly not what we want to do, since first we 
include the $\Phi_1$ which should be invisible and second the electrons 
produced in the decay chains, potentially also with high $p_T$, are 
included.

For the purpose of this tutorial we use a rather ugly trick in order 
to eliminate particles with respect to the jet reconstruction. We 
simply loop through the event record and when we find a $\Phi_1$ 
(\texttt{id == 9000006}) we set its status to a negative value, since 
the jet reconstruction only considers final state particles, 
\textit{i.e.} with positive status. In order to do this, insert the 
following lines, just before the jet analysis part,
\begin{verbatim}
   for (int iPart = 0; iPart < pythia.event.size(); ++iPart) {
      if (pythia.event[iPart].idAbs() == 9000006) { 
        int stat = pythia.event[iPart].status();
        if (stat > 0) pythia.event[iPart].status(-stat);
      }
   }
\end{verbatim}
These lines hence illustrate both how to loop over the particles in the 
event record and how to access their properties, in this case the id 
and status codes. Now compile and run the program again. The mean number 
of jets should now have been reduced by approximately two (from 4.8 to 
3.1), due to the two $\Phi_1$ in the events. The leading $p_T$ jet 
should now dominantly originate from the quarks in the $U$ decay and the 
mean $p_T$ therefore is around half of the $U$ mass, 500 GeV.

We now leave it as an exercise to also eliminate any final state 
electrons in the event and see how the distributions are affected. 
Another easy thing to investigate would be to see how the number 
of reconstructed jets varies with the minimum $p_T$ requirement 
specified for the jet finder.

\subsection{The event record}

The event record is set up to store every step in the evolution from 
an initial low-multiplicity partonic process to a final high-multiplicity
hadronic state, in the order that new particles are generated. The record
is a vector of particles, that expands to fit the needs of the current 
event (plus some additional pieces of information not discussed here). 
Thus \texttt{event[i]} is the \texttt{i}'th particle of the current 
event, and you may study its properties by using various 
\texttt{event[i].method()} possibilities.

The \texttt{event.list()} listing provides the main properties of 
each particles, by column:
\begin{Itemize}
\item \texttt{no}, the index number of the particle (\texttt{i}
above);
\item \texttt{id}, the PDG particle identity code 
(method \texttt{id()});
\item \texttt{name}, a plaintext rendering of the particle name 
(method \texttt{name()}), within brackets for initial or intermediate 
particles and without for final-state ones;
\item \texttt{status}, the reason why a new particle was added to 
the event record (method \texttt{status()});
\item \texttt{mothers} and \texttt{daughters}, documentation on
the event history (methods \texttt{mother1()}, \texttt{mother2()}, 
\texttt{daughter1()} and \texttt{daughter2()});
\item \texttt{colours}, the colour flow of the process (methods 
\texttt{col()} and \texttt{acol()});
\item \texttt{p\_x}, \texttt{p\_y}, \texttt{p\_z} and \texttt{e}, 
the components of the momentum four-vector $(p_x, p_y, p_z, E)$,
in units of GeV with $c = 1$ (methods \texttt{px()}, \texttt{py()}, 
\texttt{pz()} and \texttt{e()});
\item \texttt{m}, the mass, in units as above (method \texttt{m()}).
\end{Itemize}
For a complete description of these and other particle properties 
(such as production and decay vertices, rapidity, $p_\perp$, etc), 
open the program's online documentation in a browser (see Section \ref{sec:PYpreworkshop},
point 6, above), scroll down to the ``Study Output'' section, and follow
the ``Particle Properties'' link in the left-hand-side menu.  For brief
summaries on the less trivial of the ones above, read on.

\subsubsection{Identity codes}

A complete specification of the PDG codes is found in the
Review of Particle Physics \cite{rpp}. An online listing is 
available from\\
\hspace*{10mm}\texttt{http://pdg.lbl.gov/2008/mcdata/mc\_particle\_id\_contents.html}

A short summary of the most common \texttt{id} codes would be\\[2mm] 
\begin{tabular}{|cc|cc|cc|cc|cc|cc|cc|}
\hline
1 & $\d$ & 11 & $\e^-$      & 21  & $\g$ & 211 & $\pi^+$ 
& 111 & $\pi^0$ & 213 & $\rho^+$ & 2112 & $\n$ \\
2 & $\u$ & 12 & $\nu_{\e}$   & 22 & $\gamma$ & 311 & $\K^0$
& 221 & $\eta$ & 313 & $\K^{*0}$ & 2212 & $\p$  \\
3 & $\s$ & 13 & $\mu^-$     & 23 & $\Z^0$  & 321 & $\K^+$
& 331 & $\eta'$ & 323 & $\K^{*+}$ & 3122 & $\Lambda^0$ \\
4 & $\c$ & 14 & $\nu_{\mu}$  & 24 & $\W^+$ & 411 & $\D^+$
& 130 & $\K^0_{\mrm{L}}$ & 113 & $\rho^0$ & 3112 & $\Sigma^-$ \\
5 & $\b$ & 15 & $\tau^-$    & 25 & $\H^0$ & 421 & $\D^0$
& 310 & $\K^0_{\mrm{S}}$ & 223 & $\omega$ & 3212 & $\Sigma^0$  \\
6 & $\t$ & 16 & $\nu_{\tau}$ &  &  & 431 & $\D_{\s}^+$
& & & 333 & $\phi$ & 3222 & $\Sigma^+$ \\
\hline
\end{tabular}\\[2mm]
Antiparticles to the above, where existing as separate entities, 
are given with a negative sign.\\
Note that simple meson and baryon codes are constructed from 
the constituent (anti)quark codes, with a final spin-state-counting digit 
$2 s + 1$ ($\K^0_{\mrm{L}}$ and $\K^0_{\mrm{S}}$ being exceptions), and with
a set of further rules to make the codes unambiguous.  

\subsubsection{Status codes}

When a new particle is added to the event record, it is assigned 
a positive status code that describes why it has been added, 
as follows:\\[2mm] 
\begin{tabular}{|c|l|}
\hline
code range & explanation \\
\hline
11 -- 19 & beam particles\\
21 -- 29 & particles of the hardest subprocess\\
31 -- 39 & particles of subsequent subprocesses in multiparton interactions\\
41 -- 49 & particles produced by initial-state-showers\\
51 -- 59 & particles produced by final-state-showers\\
61 -- 69 & particles produced by beam-remnant treatment\\
71 -- 79 & partons in preparation of hadronization process\\
81 -- 89 & primary hadrons produced by hadronization process\\
91 -- 99 & particles produced in decay process, or by Bose-Einstein effects\\ 
\hline
\end{tabular}\\[2mm]
Whenever a particle is allowed to branch or decay further its status 
code is negated (but it is \textit{never} removed from the event record), 
such that only particles in the final state remain with positive codes. The
\texttt{isFinal()} method returns \texttt{true/false} for
positive/negative status codes.

\subsubsection{History information}

The two mother and two daughter indices of each particle provide 
information on the history relationship between the different entries 
in the event record. The detailed rules depend on the particular physics 
step being described, as defined by the status code. As an example, 
in a $2 \to 2$ process $a b \to c d$, the locations of $a$ and $b$
would set the mothers of $c$ and $d$, with the reverse relationship
for daughters. When the two mother or daughter indices are not
consecutive they define a range between the first and last entry,
such as a string system consisting of several partons fragment into
several hadrons.

There are also several special cases. One such is when ``the same''
particle appears as a second copy, e.g. because its momentum has 
been shifted by it taking a recoil in the dipole picture of parton
showers. Then the original has both daughter indices pointing to the
same particle, which in its turn has both mother pointers referring
back to the original. Another special case is the description of 
ISR by backwards evolution, where the mother is constructed at a 
later stage than the daughter, and therefore appears below in the 
event listing. 

If you get confused by the different special-case storage options, the 
two \texttt{pythia.event.motherList(i)} and
\texttt{pythia.event.daughterList(i)} methods are able to return a
\texttt{vector} of all mother or daughter indices of particle
\texttt{i}.

\subsubsection{Color flow information}

The colour flow information is based on the Les Houches Accord
convention \cite{leshouchesaccord}. In it, the number of colours
is assumed infinite, so that each new colour line can be assigned
a new separate colour. These colours are given consecutive labels:
101, 102, 103, \ldots . A gluon has both a colour and an anticolour
label, an (anti)quark only (anti)colour. 

While colours are traced consistently through hard processes and 
parton showers, the subsequent beam-remnant-handling step often 
involves a drastic change of colour labels. Firstly, previously 
unrelated colours and anticolours taken from the beams may at this
stage be associated with each other, and be relabeled accordingly. 
Secondly, it appears that the close space--time overlap of many 
colour fields leads to reconnections, i.e. a swapping of colour labels, 
that tends to reduce the total length of field lines.

\subsection{The jet analysis program}

This is the final \texttt{mymain.cc} at the end of the tutorial,
\begin{verbatim}
#include "Pythia.h" // Include Pythia headers.
using namespace Pythia8; // Let Pythia8:: be implicit.

int main() { // Begin main program.

  // Set up generation.
  Pythia pythia; // Declare Pythia object
  pythia.readString("Beams:frameType = 4"); // Beam info in LHEF.
  pythia.readString("Beams:LHEF = events_with_decay.lhe"); 
  pythia.init(); // Initialize; incoming pp beams is default.

  SlowJet::SlowJet sJets(-1, 0.4, 50., 5., 1, 2);
  Hist nJ( "Nr Jets", 10, -0.5, 9.5);
  Hist j1pT( "Leading Jet pT", 100, 0., 1000.);

  // Generate event(s).
  for (int iEvent = 0; iEvent < 1000; ++iEvent) {
    if (!pythia.next()) {
      if (pythia.info.atEndOfFile()) break; // Exit at enf of LHEF.
      continue; // Skip event in case of problem.
    }

    // Loop through event record.
    for (int iPart = 0; iPart < pythia.event.size(); ++iPart) {
      if (pythia.event[iPart].idAbs() == 9000006) { 
        int stat = pythia.event[iPart].status();
        if (stat > 0) pythia.event[iPart].status(-stat);
      }
      if (pythia.event[iPart].idAbs() == 11) { 
        int stat = pythia.event[iPart].status();
        if (stat > 0) pythia.event[iPart].status(-stat);
      }
    }

    // Jet analysis.
    if (sJets.analyze(pythia.event)) {
      nJ.fill(sJets.sizeJet());
      j1pT.fill(sJets.pT(0));
    }


  }
  pythia.stat(); // Print run statistics.

  std::cout << nJ << j1pT;

  return 0;
} 
\end{verbatim}

\section{\Herwigpp tutorial}
\label{sec:HW}

\subsection {Introduction}
\Herwigpp~\citep{Bahr:2008pv, Gieseke:2011na}, is a general-purpose Monte Carlo event generator for the
simulation of hard lepton-lepton, lepton-hadron and hadron-hadron
collisions. 

Some of the main features, in random order, are:
\begin{enumerate}
\item{Many matrix elements natively and capability of reading in Les
    Houches-accord event files.}
\item{A description of the underlying event via multiple parton
    interactions.} 
\item{Initial- and final-state QCD parton showering.}
\item{Next-to-leading order: Several {\sc Powheg}~\citep{Nason:2006hfa,
      LatundeDada:2006gx} processes and
    compatibility with {\sc MC@NLO} \citep{Frixione:2002bd}.}
\item{Cluster hadronization.}
\item{Individually modelled hadron and tau decays.}
\item{QED radiation.}
\item{Analysis via {\sc Rivet}~\citep{Buckley:2010ar}, native analysis handlers, or output
    to \texttt{HepMC} events.}
\end{enumerate}

This tutorial aims to guide you through the steps of going from the \Herwigpp source distribution
through to some basic analysis for a specific model, which is
described in Section~\ref{sec:model}. The model events are generated in
{\sc MadGraph}~\citep{Alwall:2011uj}, and will be `fed' into the
event generator via the Les Houches interface\footnote{In the future,
  \Herwigpp will include automated creation of the model internally
  via a {\sc FeynRules} UFO interface~\citep{Degrande:2011ua}.}. 

Note that Wiki page for this tutorial also exists at:\\
\url{http://herwig.hepforge.org/trac/wiki/MC4BSM}.

\subsection{Installation and setup}
\subsubsection{Required libraries}
Throughout this tutorial, it will be assumed that you have created a
working directory in your home directory, which we will be
referring to as \verb+$MCBSM+ from now on. If you wish, you may
define \verb+$MCBSM+ (in bash shell):
\begin{verbatim}
export MCBSM=/target/directory
\end{verbatim}
In c-shell you need to use:
\begin{verbatim}
setenv MCBSM "/target/directory" 
\end{verbatim}
All the source tarballs will be assumed to be placed in
\verb+$MCBSM/source+. 

If the libraries do not already exist on your system, you will need to compile
and install the following:
\begin{itemize}
\item{GNU Scientific Library (\texttt{GSL}),\\
    \url{http://mirror.switch.ch/ftp/mirror/gnu/gsl/gsl-1.15.tar.gz}.\\
    Recommended installation in \verb+$MCBSM/gsl+.}
\item{{\sc LHAPDF}, the Les Houches Accord PDF Interface,\\
    \url{http://www.hepforge.org/archive/lhapdf/lhapdf-5.8.7.tar.gz}. \\
    Recommended installation in \verb+$MCBSM/lhapdf+.} 
\item{The PDF used in this tutorial, cteq6ll,\\
    \url{http://www.hepforge.org/archive/lhapdf/pdfsets/current/cteq6ll.LHpdf}. \\
    To be placed in \verb+$MCBSM/lhapdf/share/lhapdf/PDFsets+ (the directory may need to be created). }
\end{itemize}

Important: Make sure that any program binaries and libraries you
install appear in your \verb+$PATH+ and \verb+$LD_LIBRARY_PATH$+
(\verb+$DYLD_LIBRARY_PATH+ in \texttt{MacOS X}) respectively,
otherwise the binaries or libraries will not be found by the system. 

Note that when compiling, if you have a multi-core machine, you may use \verb+make -jX+, where \verb+X+ is the number of cores, to speed up the process.

\paragraph{\texttt{GSL}}
Compilation and installation in the conventional way:
\begin{verbatim}
tar xvzf gsl-1.15.tar.gz; cd gsl-1.15
./configure --prefix=$MCBSM/gsl
make
\end{verbatim}
If everything is successful (i.e. there are no errors):
\begin{verbatim}
make install
\end{verbatim}
\paragraph{\texttt{LHAPDF}}
Equivalent to \texttt{GSL} with the necessary change of target
directory through:
\begin{verbatim}
--prefix=$MCBSM/lhapdf
\end{verbatim}
Once installed, make sure cteq6ll.LHpdf is placed in \verb+$MCBSM/lhapdf/share/lhapdf/PDFsets+. 

\subsubsection{Installing \Herwigpp and {\sc ThePEG}}
\Herwigpp is based on {\sc ThePEG}, (Toolkit for High Energy
Physics Event Generation), which is essentially a library that
contains the necessary building blocks for constructing an event
generator. 

First we need to compile and install \textsc{ThePEG}. Get the latest version at:\\
\url{http://www.hepforge.org/archive/thepeg/ThePEG-1.7.3.tar.bz2}.\\
To compile and install:
\begin{verbatim}
tar xvfj ThePEG-1.7.3.tar.bz2; cd ThePEG-1.7.3
./configure --prefix=$MCBSM/thepeg --with-gsl=$MCBSM/gsl 
--with-LHAPDF=$MCBSM/lhapdf
make; make install
\end{verbatim}
For instructions on how to do this on the MacOS X operating system,
consult:\\
\url{http://herwig.hepforge.org/trac/wiki/HwOsX}\\

Once \textsc{ThePEG} is installed, you are ready to install \Herwigpp,
linking to all the necessary libraries. Get the latest stable release
(currently version 2.5.2) at:\\
\url{http://www.hepforge.org/archive/herwig/Herwig++-2.5.2.tar.bz2}.\\
To compile and install:
\begin{verbatim}
tar xvfj Herwig++-2.5.2.tar.bz2; cd Herwig++-2.5.2
./configure --prefix=$MCBSM/hpp --with-thepeg=$MCBSM/thepeg
--with-gsl=$MCBSM/gsl 
make; make install
\end{verbatim}

\subsubsection{\Herwigpp preliminaries}
To run \Herwigpp one needs to set up a working directory. In this
tutorial, this will be taken to be \verb+$MCBSM/hppwork+. Create this
directory and copy the following:\footnote{The subdirectory \texttt{source} will not be required if you have not put the source files in there!}
\begin{verbatim}
cp $MCBSM/source/Herwig++-2.5.2/src/*.in $MCBSM/hppwork
cp $MCBSM/source/Herwig++-2.5.2/src/*.model $MCBSM/hppwork
cp $MCBSM/source/Herwig++-2.5.2/src/defaults/*.in $MCBSM/hppwork
\end{verbatim}
This copies all the necessary input files, including the ones
containing the default settings, as well as different \verb+.model+ files (such as
the MSSM, ADD, and so on) for future use. Once this is done, change into the
\verb+$MCBSM/hppwork+ directory and execute:
\begin{verbatim}
Herwig++ init
\end{verbatim}
If the program has compiled correctly, the binaries and libraries are
in your \verb+$PATH+ and \verb+$LD_LIBRARY_PATH+ respectively, and all
the necessary \verb+.in+ files have been copied properly, there should be no
errors.

Most of the copied \verb+.in+ files contain inputs for the default
settings (e.g. \verb+PDF.in+ includes defaults for PDF handling,
\verb+Cuts.in+ includes the default cuts and so on). It is
not recommended in general to change the settings for running in those
files directly. Instead, one should use a customized input
file for the individual runs. The simplest example to look at is the file \verb+LHC.in+. In
there, you need to choose the desired energy at which you wish
to run the event generator. The default in this input file is LHC at 7 TeV:
\begin{verbatim}
########################
## sqrt(s) = 7000 GeV ##
########################
set LHCGenerator:EventHandler:LuminosityFunction:Energy 7000.0
set /Herwig/UnderlyingEvent/KtCut:MinKT 3.06
set /Herwig/UnderlyingEvent/UECuts:MHatMin 6.12
\end{verbatim}
You may wish to examine the construction of the \verb+LHCGenerator+ in
the file \verb+HerwigDefaults.in+, which is of type
\verb+EventHandler+.\footnote{The \texttt{EventHandler} class is a
 general class of \textsc{ThePEG}, responsible for the generation of events.} . It is recommended \textit{not} to make any
changes in \verb+HerwigDefaults.in+, instead, these should be made in
\verb+LHC.in+, as also mentioned previously. 
A useful thing to be able to do is change the type of the hard process
to be generated. The default in \verb+LHC.in+ is Drell-Yan Z/photon:
\begin{verbatim}
# Drell-Yan Z/gamma
insert SimpleQCD:MatrixElements[0] MEqq2gZ2ff
\end{verbatim}
Let's comment this out (by adding a \verb+#+ in front) and instead uncomment: 
\begin{verbatim}
# top-antitop production
insert SimpleQCD:MatrixElements[0] MEHeavyQuark
\end{verbatim}
and execute:
\begin{verbatim}
Herwig++ read LHC.in
\end{verbatim}
If there are no errors, the command should return no text output. It
should create a file \verb+LHC.run+, which will contain internal
instructions to the event generator, which are not human-readable. To make a
short run, say, of 100 events of top-antitop events, including initial/final-state radiation, hadronization, and the underlying
event, simply type:
\begin{verbatim}
Herwig++ run LHC.run -N100
\end{verbatim}
The run will initialize and commence generating events. \textbf{This point is
a good place to stop the pre-workshop part of the tutorial.}

Running the file with the above command will not produce
any significant output apart from the file \verb+LHC.out+, which will
contain the cross section of the chosen process (you may even choose
more than one process). A log file may
obtained by adding \verb+-d1+ to the above:
\begin{verbatim}
Herwig++ run LHC.run -N100 -d1
\end{verbatim}
This will print the first 10 events of the run (determined by
\verb+setLHCGenerator:PrintEvent 10+ in the \verb+LHC.in+ file) in
\verb+LHC.log+. Have a look at it and see if you can recognize the
different steps of the event generation, and follow through the decay
products of the top and antitop. For a full explanation have a look at
the Wiki page:\\
\url{http://herwig.hepforge.org/trac/wiki/EventRecordFormat}.\\
You may also wish to enable the built-in analysis for top-antitop:
\begin{verbatim}
# analysis of top-antitop events
insert LHCGenerator:AnalysisHandlers 0 /Herwig/Analysis/TTbar
\end{verbatim}
Note that every time a change is made in the \verb+.in+ file, one needs to
invoke the `read' command. 

After running the event generator again, the analysis will produce a file
\verb+LHC-TTbar.top+. This is made to be read by \textsc{TopDrawer}, a rather
archaic, but useful, plotting package. Unfortunately it is difficult
to compile it from source, and pre-compiled binaries may or may not
work on your system. You may try the binaries subdirectory of:\\
\url{http://www.rcnp.osaka-u.ac.jp/~okamura/ftp/pub/index.php?dir=topdrawer/}\\
If the executable works, you may try to create a postscript file by
running:\\
\verb+td -d ps LHC-TTbar+,\\
and using \texttt{GhostView} (\texttt{gv}) to view it. 

To run through the rest of the tutorial, you will need to obtain the following
tarball:\\
\url{http://www.itp.uzh.ch/~andreasp/mc4bsm.tar.gz}\\
Extracting this in \verb+$MCBSM+ should create a subdirectory
mc4bsm. The details of the contents of \verb+$MCBSM+ will be discussed in Section~\ref{bsmtutorial}. 

\subsection{The on-site tutorial}

This tutorial will study the toy model described in Section~\ref{sec:model}. 
We assume the existence of {\sc MadGraph} parton-level (but with the heavy new particles
already decayed) event files (see Section~\ref{sec:MG}). 

\subsubsection{Running \Herwigpp  with the toy model}\label{bsmtutorial}

Extracting \verb+mc4bsm.tar.gz+ in \verb+$MCBSM+ should create a subdirectory
\verb+mc4bsm+. This in turn contains the following subdirectories:
\begin{itemize}
\item{\verb+analysis+: contains the sources of a basic \Herwigpp
    analysis.}
\item{\verb+hppinput+: contains the input files necessary to run the 
    basic tutorial events.}
\item{\verb+madgraph_events+: contains the {\sc MadGraph} events
    which include decays of the new heavy particles down to `stable' particles.}
\item{\verb+results+: contains `solution' plots to the tutorials.}
\end{itemize}

We first need to compile the customized analysis handler. To do that,
copy the \verb+MC4BSMAnalysis.tar.gz+ to the \verb&Herwig++-2.5.2/Contrib&
subdirectory, and extract:
\begin{verbatim}
tar xvzf MC4BSMAnalysis.tar.gz 
\end{verbatim}
This should overwrite the default
\verb&Herwig++-2.5.2/Contrib/Makefile.am& and put the analysis source
code in the \verb+MC4BSMAnalysis+. In the \verb&Herwig++-2.5.2/Contrib&
directory, type:
\begin{verbatim}
sh make_makefiles.sh
\end{verbatim}
This forces the recreation of all the \verb+Makefile+s from input, and
hence also creates a \verb+Makefile+ for our analysis. Change to the
\verb&Herwig++-2.5.2/Contrib/MC4BSMAnalysis& directory and type
\verb+make+. This should compile the library
\verb+MC4BSMAnalysis.so+. Copy this into the \Herwigpp installation
directory library:
\begin{verbatim}
cp MC4BSMAnalysis.so $MCBSM/hpp/lib/Herwig++
\end{verbatim}

In your \Herwigpp working directory (\verb+$MCBSM/hppwork+), copy the
contents of \verb+hppinput+ and \verb+madgraph_events+. These should be the input file
\verb+MC4BSM.in+ and the model file \verb+MC4BSM.model+ from the prior
directory and \verb+events_with_decays.lhe+ from the latter. Have a look
and try to understand at what's in each of the \verb+MC4BSM.*+ files:
The \verb+MC4BSM.in+ file is equivalent to the \verb+LHC.in+ input file, with the
appropriate settings to read in a Les Houches-accord event file. The
underlying event settings have been chosen for an anticipated 8~TeV
run. There's also a change of PDF to the one used by the input events
(cteq6ll), using the \texttt{LHAPDF} library. Even though the input event file provides all the necessary
information for the hard process, \Herwigpp needs to know about the
existence of the new particles: that is why the second file,
\verb+MC4BSM.model+, is necessary, and is read at the top of \verb+MC4BSM.in+. It's easy to construct one
manually:\footnote{In a future release, it will be possible to construct
  it using the {\sc FeynRules} interface.} for example, the following defines particle \verb+uv+ with PDG id
9000008, mass $500\gev$ and width $1\gev$. 
\begin{verbatim}
create ThePEG::ParticleData uv
setup uv 9000008 uv 500.0 1.0 10.0 1.973269631e-13 2 3 2 0
\end{verbatim}
The rest of the numbers in order are: maximum number of widths away
from central value of mass allowed,  the lifetime (calculated from the
width), the electromagnetic charge (units of 1/3), the $SU(3)_c$
charge (3 implies triplet, 1 implies singlet and so on), the spin $2s+1$ and a boolean for whether
the particle is stable or not. 

Read in and run the input file in the usual way:
\begin{verbatim}
Herwig++ read MC4BSM.in
Herwig++ run MC4BSM.run -N9000
\end{verbatim}
The analysis should create a subdirectory \verb+mcplots+ containing
some plots, in \texttt{TopDrawer} format. You may view them using TopDrawer, or
copy the data in your favourite plotting program. The plot files are
the pseudorapidities and transverse momenta of the first 6
highest-$p_T$ jets in the event (where in this case we are also
including the invisibles), the number of jets (\verb+Njets+) and the
total invariant mass in the event (including the invisibles),
calculated by:
\beq
M^2 = E^2 - \vec{P}^2\;\;,
\eeq
where $E$ and $\vec{P}$ are the sums of the all the individual particle
energies and 3-momenta respectively, within a certain range of
pseudorapidity. The pseudorapidity cut for the jets, as well as other options for the
analysis can be found at the relevant section in the input file:
\begin{verbatim}
# Choose jet clustering algorithm
set /Herwig/Analysis/MC4BSMAnalysis:JetAlgorithm 2 
# Jet ET cut to apply in jet clustering 
set /Herwig/Analysis/MC4BSMAnalysis:ETClus 20*GeV
# Cone size used in clustering in merging.
set /Herwig/Analysis/MC4BSMAnalysis:RClus 0.4
# Max |eta| for jets in clustering in merging.
set /Herwig/Analysis/MC4BSMAnalysis:EtaClusMax 5
\end{verbatim}
The options \verb+ETClus+, \verb+RClus+ and \verb+EtaClusMax+ correspond to the
minimum $p_T$ of the jets, the clustering parameter for the algorithm
and the maximum pseudorapidity for the jets.

Try and see what happens when you switch off the hadronization and
underlying event, by uncommenting the relevant pieces:
\begin{verbatim}
#set LesHouchesHandler:HadronizationHandler  NULL              
#set /Herwig/Shower/ShowerHandler:MPIHandler NULL    
\end{verbatim}

You may wish to edit the analysis code and see if you can add an extra
plot. This can be done by editing \verb+MC4BSMAnalysis.cc+ and
\verb+MC4BSMAnalysis.h+ in the
\verb%Contrib/MC4BSMAnalysis% subdirectory. You will
     need to add a pointer for the new histogram in the \verb+.h+ file
     and then add the analysis code in the \verb+.cc+ file to declare
     the limits of the histogram, fill it in, and write it out. Take
     the existing histograms as an example. A good one to try is the
     $p_T$ of all leptons, for which there already exists
     commented-out code at the right places in the code for guidance:
     search for ``\verb+EXAMPLE NEW HISTOGRAM:+'' in the code. There
     should be three points in the \verb+.cc+ file and one in the
     \verb+.h+ file. 

\subsubsection{Further reading}
For specifics of the \Herwigpp event generator, the manual and release
notes for version 2.5 are useful~\citep{Bahr:2008pv,
  Gieseke:2011na}. For further, up-to-date information, you may
consult the Wiki pages, at:\\
\url{http://herwig.hepforge.org/trac/}. 

For details on the specific inner workings of \Herwigpp and {\sc ThePEG},
consult the \texttt{Doxygen} output at:\\
\url{http://herwig.hepforge.org/doxygen/index.html}\\
\url{http://thepeg.hepforge.org/doxygen/index.html}\\
Although it may look a bit difficult to navigate through, it is a good
starting point to find out how things are interconnected inside the event
generator.

\section{{\sc Sherpa} tutorial}
\label{sec:SH}

\subsection{Requirements}
You will need a PC running Linux or a Mac to follow this exercise.
Many of the Macs the {\sc Sherpa} team came across required some special
tweaks to guarantee smooth installation. Linux should therefore be
the preferred operating system for this exercise.

You should have the GNU autotools installed.
They will also come in handy for other applications.
Try to locate \verb!libtoolize!, \verb!automake! and \verb!autoconf!
on your system. If they are missing, and there is no package manager
that provides easy installation, you can download the sources from
\begin{itemize}
\item \href{http://ftp.gnu.org/gnu/autoconf/autoconf-2.65.tar.gz}{
            http://ftp.gnu.org/gnu/autoconf/autoconf-2.65.tar.gz}
\item \href{http://ftp.gnu.org/gnu/automake/automake-1.11.1.tar.gz}{
            http://ftp.gnu.org/gnu/automake/automake-1.11.1.tar.gz}
\item \href{http://ftp.gnu.org/gnu/libtool/libtool-2.2.6b.tar.gz}{
            http://ftp.gnu.org/gnu/libtool/libtool-2.2.6b.tar.gz}
\end{itemize}
Follow the installation instructions provided in the tarballs.

Although not mandatory for this exercise, it will be useful to have
metapost installed. It will be used to plot the Feynman graphs generated by {\sc Sherpa}.

To visualize the analysis results, you will need gnuplot or a similar plotting tool.
The instructions given here assume that you have gnuplot installed.
\subsection{Installation}
There are two methods to install {\sc Sherpa} on your system
\subsubsection{... via package download}
The {\sc Sherpa} sources for this exercise are obtained from either of these sites
\begin{itemize}
\item \href{http://sherpa.hepforge.org/trac/wiki/SherpaDownloads/Sherpa-1.4.0}{
            http://sherpa.hepforge.org/trac/wiki/SherpaDownloads/Sherpa-1.4.0}
\item \href{http://www.slac.stanford.edu/~shoeche/mc4bsm/Sherpa-1.4.0}{
            http://www.slac.stanford.edu/$\sim$shoeche/mc4bsm/Sherpa-1.4.0}
\end{itemize}
After the package download go to the download directory and follow these steps
\begin{verbatim}
  tar -xzf SHERPA-MC-1.4.0.tar.gz
  cd SHERPA-MC-1.4.0/
  ./configure --enable-analysis
  make -j2 install
\end{verbatim}
If successful, you should now find a {\sc Sherpa} executable in the \verb!bin/! subdirectory.
On a Mac with 64 bit system, use \verb!FC='gfortran -m64' ./configure --enable-analysis! 
instead of the above configure command. 
\subsubsection{... via svn checkout}
If you have a subversion client installed on your system, you can check out {\sc Sherpa}
as follows. This will be an easier option if you try to keep your version
up-to-date with bugfixes. Note that you {\em must} have the GNU autotools installed 
to proceed with this method. Follow these steps
\begin{verbatim}
  svn co http://sherpa.hepforge.org/svn/branches/rel-1-4-0
  cd rel-1-4-0/
  autoreconf -i
  ./configure --enable-analysis
  make -j2 install
\end{verbatim}
If successful, you should now find a {\sc Sherpa} executable in the \verb!bin/! subdirectory.
On a Mac with 64 bit system, use \verb!FC='gfortran -m64' ./configure --enable-analysis! 
instead of the above configure command. 
\subsubsection{Dealing with problems}
If there are problems during the installation process, please send an email to
\href{mailto:info@sherpa-mc.de}{info@sherpa-mc.de} or list the issue on our
\href{http://sherpa.hepforge.org/trac/report}{bug tracker}.\
Some advice can also be found in Section 2 - ``Getting started'' - 
of the online manual, which is available at
\href{http://sherpa.hepforge.org/doc/SHERPA-MC-1.4.0.html}{
      http://sherpa.hepforge.org/doc/SHERPA-MC-1.4.0.html}
\subsection{Testing your {\sc Sherpa} installation}
To guarantee a successful installation of {\sc Sherpa}, we will try to simulate 
the production of $W^\pm$-boson plus jets final states at the Tevatron
using ME+PS merging with up to two jets only (for simplicity). Follow these steps
\begin{verbatim}
  cd Examples/Tevatron_WJets/
  ../../bin/Sherpa
\end{verbatim}
The example should take about 15 minutes to run.\\
You should see the following output
\begin{verbatim}
Welcome to Sherpa, <username>. Initialization of framework underway.
Run_Parameter::Init(): Setting memory limit to <limit> GB.
-----------------------------------------------------------------------------
-----------    Event generation run with SHERPA started .......   -----------
-----------------------------------------------------------------------------
\end{verbatim}
... lots more following ...
\begin{verbatim}
+-----------------------------------------------------+
|                                                     |
|  Total XS is 2338.05 pb +- ( 22.9361 pb = 0.98 % )  |
|                                                     |
+-----------------------------------------------------+
\end{verbatim}
... some more following ...
\begin{verbatim}
Time: 17m 51s on Thu Mar  8 12:31:59 2012
 (User: 17m 49s, System: 2s, Children User: 0s, Children System: 0s)
\end{verbatim}
If this output shows up in your terminal, you have successfully
installed {\sc Sherpa}.\\

{\large Congratulations!}

\subsubsection{Dealing with problems}
If {\sc Sherpa} refuses to run, please send an email to
\href{mailto:info@sherpa-mc.de}{info@sherpa-mc.de} or list the issue on our
\href{http://sherpa.hepforge.org/trac/report}{bug tracker}.\

\subsection{Getting started}
The setup is available from
\begin{itemize}
\item \href{http://www.slac.stanford.edu/~shoeche/mc4bsm/2012/}{
            http://www.slac.stanford.edu/$\sim$shoeche/mc4bsm/2012/}
\end{itemize}
Download the tarball \verb!mc4bsm_sherpa.tar.gz!
and unpack the setup files using
\begin{verbatim}
  tar -xzf mc4bsm_sherpa.tar.gz
\end{verbatim}

Note that we use a given set of {\sc FeynRules} output files, which are stored
in the subdirectory \verb!FeynRules_Output!. You can overwrite these files
with your own. However, if you alter the masses of $U$, $E$, $\phi_1$ or $\phi_2$,
the particle widths have to be recomputed. Section~\ref{sec:widths} explains how to do this.
You can work through Sec.~\ref{sec:widths} first, but it is recommended to follow
Secs.~\ref{sec:intro}-\ref{sec:widths} in order, as this will help
to understand the input structure of {\sc Sherpa}.

For the remainder of this exercise we will assume that {\sc Sherpa} was 
installed into a directory called \verb!<sherpa_prefix>/!. 
You should be able to locate the executeable in the subdirectory 
\verb!<sherpa_prefix>/bin/!, cf.\ the pre-workshop instructions.

\subsection{Understanding the input structure}
\label{sec:intro}
Change to the directory \verb!Intro! and open the input file 
\verb!Run.dat!. The file consists of various sections, which are marked 
as \verb!(model)!, \verb!(beam)!, \verb!(process)!, etc.

The \verb!(model)! section contains an instruction to use the 
{\sc Sherpa}-internal {\sc FeynRules} interface and to flag the particles $E$, 
$\phi_1$ and $\phi_2$ as unstable.

The \verb!(beam)! section is used to set up the collider type ($pp$)
and its cms energy (8~TeV).

The \verb!(processes)! section defines the reaction of interest.
Here we simulate the process $u\bar{u}\to U\bar{U}$.
We also instruct {\sc Sherpa} to write out Latex files depicting the
Feynman graphs.

The \verb!(me)! section is used to set a scale 
at which the strong coupling is to be evaluated.

Run the simple example using
\begin{verbatim}
  <sherpa_prefix>/bin/Sherpa
\end{verbatim}
{\sc Sherpa} will stop with the message
\begin{verbatim}
   New libraries created. Please compile.
\end{verbatim}
followed by some citation info. Now you need to compile 
the process-specific source code generated by {\sc Sherpa} using
\begin{verbatim}
  ./makelibs
\end{verbatim}
After the compilation has finished, run {\sc Sherpa} again
\begin{verbatim}
  <sherpa_prefix>/bin/Sherpa
\end{verbatim}
The program will now dynamically link the libraries which
have just been created and compute a cross section for the
process $u\bar{u}\to U\bar{U}$.\\
Have a look at the Feynman graphs that contribute to this process
\begin{verbatim}
  ./plot_graphs Process/P2_2/
\end{verbatim}

\subsection{Simulating parton-level events}
Change back to the original directory and then
go to \verb!ToyModel_PartonLevel!.

Have a look at the \verb!Run.dat! input file.

In the \verb!(run)! section we disable the hadronzation module 
by using \verb!FRAGMENTATION Off!, the parton shower by using 
\verb!SHOWER_GENERATOR None! and the YFS soft photon generator 
by using \verb!ME_QED Off!. 

In the \verb!(processes)! section, we instruct {\sc Sherpa} to generate
the following reactions:
\begin{itemize}
\item $pp\to U[\to\phi_1 u]\;\;\bar{U}[\to\phi_1 \bar{u}]$
\item $pp\to U[\to\phi_2[\to\phi_1\; e^+e^-]\; u]\;\;\bar{U}[\to\phi_1 \bar{u}]$
\item $pp\to U[\to\phi_1 u]\;\;\bar{U}[\to\phi_2[\to\phi_1\; e^+e^-]\; \bar{u}]$
\item $pp\to U[\to\phi_2[\to\phi_1\; e^+e^-]\; u]\;\;\bar{U}[\to\phi_2[\to\phi_1\; e^+e^-]\; \bar{u}]$
\end{itemize}

Run this setup using
\begin{verbatim}
  <sherpa_prefix>/bin/Sherpa
\end{verbatim}
Again, {\sc Sherpa} will stop with the message
\begin{verbatim}
   New libraries created. Please compile.
\end{verbatim}
followed by some citation info. You need to compile 
the process-specific source code generated by {\sc Sherpa} 
using \verb!./makelibs!.
After the compilation has finished, run {\sc Sherpa} again
\begin{verbatim}
  <sherpa_prefix>/bin/Sherpa
\end{verbatim}
The cross sections will be computed and {\sc Sherpa} will generate 10000 events.
Near the end of the output you should see the following message
\begin{verbatim}
+---------------------------------------------------------+
|                                                         |
|  Total XS is 0.444465 pb +- ( 0.00429378 pb = 0.96 % )  |
|                                                         |
+---------------------------------------------------------+
\end{verbatim}
where the precise value of the cross section depends on the mass parameters
you have chosen for the $U$, $E$ and $\phi_{1/2}$ fields.
It is important that you use this cross section, when computing
event rates, rather than any of the cross sections quoted during 
the integration step. A detailed explanation why can be found in 
section three of the {\sc Sherpa} online manual,\\
\href{http://sherpa.hepforge.org/doc/SHERPA-MC-1.4.0.html#Cross-section}{
  http://sherpa.hepforge.org/doc/SHERPA-MC-1.4.0.html\#Cross-section}.

\subsection{Simulating and analyzing hadron-level events}
We are now in place to generate full events and analyze them
with {\sc Sherpa}. For simplicity, we will not include a detector simulation
in this exercise. However, {\sc Sherpa} can be combined e.g.\ with PGS to simulate
detector effects. For details on this procedure, please refer to the 
online manual, section 8.5\\
\href{http://sherpa.hepforge.org/doc/SHERPA-MC-1.4.0.html#PGS-interface}{
  http://sherpa.hepforge.org/doc/SHERPA-MC-1.4.0.html\#PGS-interface}.
{\sc Sherpa} also has a built-in Rivet-interface and it can output events
in HepMC format. For more details, please refer to the online manual.

Change back to the original directory and then
go to \verb!ToyModel_HadronLevel!.
Have a look at the \verb!Run.dat! input file.

In the \verb!(run)! section we have removed the switches that disabled
parton showers and fragmentation. Instead there is a new switch, instructing
{\sc Sherpa} to perform a simple analysis.

Open the \verb!Analysis.dat! input file. It contains instructions for the
built-in analysis module. 
\begin{itemize}
\item \verb!Finder 93 20 -4.5 4.5 0.4 1!\\
  Construct $k_T$-jets (kf-code 93) with $D=0.4$, $p_T>20$ GeV and $|\eta|<4.5$.
\item \verb!Finder 11 15 -2.5 2.5!\\
  Reconstruct electrons (kf-code 11) with $p_T>15$ GeV and $|\eta|<2.5$.
\item \verb!DRMin 11 -11 0.2!\\
  Require electrons to be separated from each other by $\Delta R>0.2$.
\item \verb!DRMin 11 93 0.4!\\
  Require electrons to be separated from jets by $\Delta R>0.4$.
\end{itemize}
Finally, we analyze the di-jet invariant mass distribution in the range
$0\le m_{jj}\le 2000$ on a linear scale with 100 bins:
\begin{verbatim}
  Mass 93 93 0 2000 100 Lin LeptonsJets
\end{verbatim}

Run this setup using
\begin{verbatim}
  <sherpa_prefix>/bin/Sherpa
\end{verbatim}
As we have linked the \verb!Process/! and \verb!Results/! directory from
the previous run, {\sc Sherpa} will immediately start generating events.

Once it has finished, plot the results of the analysis using
\begin{verbatim}
  ./plot_results.sh
\end{verbatim}
and have a look at the invariant mass distributions in \verb!plots.ps!.
Page one shows the changes when going from a pure parton-level simulation
to hadron level, while page two has separate analyses of the various
sub-processes.

\subsection{Computing the partial widths}
\label{sec:widths}
Change to the directory \verb!Widths! and open the input file 
\verb!Run.dat!.\\ The \verb!(processes)! section contains setups
for the following decays
\begin{itemize}
\item $U\to u\, \phi_1$
\item $U\to u\, \phi_2$
\item $E\to e^- \phi_1$
\item $\phi_2\to e^- e^+ \phi_1$
\end{itemize}
Comix is activated in the \verb!(me)! section using \verb!COMIX_ALLOW_BSM 1!.

Let {\sc Sherpa} compute the widths using
\begin{verbatim}
  <sherpa_prefix>/bin/Sherpa
\end{verbatim}
Update the widths in the file \verb!Particle.dat!. This file is linked
from \verb!../FeynRules_Output! and will be used in all toy model setups.

\subsection{Z+j backgrounds at LO and at NLO}
We will now proceed to simulate some important Standard-Model backgrounds.
{\sc Sherpa} will run substantially longer than in the previous steps.
You may consider doing the following part of the tutorial in parallel 
with another tutorial or after the workshop.

Change back to the original directory and then
go to \verb!Backgrounds_ZJets!.
Have a look at the \verb!Run.dat! input file.

In the \verb!(processes)! section you find the following:
\begin{verbatim}
  Process 93 93 -> 11 -11 93{2}
  Order_EW 2; CKKW sqr(30/E_CMS);
  End process;
\end{verbatim}
These lines instruct {\sc Sherpa} to generate the process $pp\to e^+e^-$
with up to two additional light partons. The tag \verb!CKKW! indicates
that the various sub-processes are to be merged using the truncated 
parton shower scheme. \verb!sqr(30/E_CMS)! sets the value of
$Q_{\rm cut}$ to 30~GeV.

Run this setup using
\begin{verbatim}
  <sherpa_prefix>/bin/Sherpa
\end{verbatim}
Once {\sc Sherpa} has finished, plot the results of the analysis using
\begin{verbatim}
  ./plot_results.sh
\end{verbatim}
and have a look at the invariant mass distributions in \verb!plots.ps!.

If you like, check out the difference between a tree-level prediction
of the background and the respective ME{\sc NLO}PS result.
Run the ME{\sc NLO}PS setup using
\begin{verbatim}
  <sherpa_prefix>/bin/Sherpa -f Run.NLO.dat
\end{verbatim}
Once {\sc Sherpa} has finished, plot the results of the analysis again, using
\begin{verbatim}
  ./plot_results.sh
\end{verbatim}
and have a look at the invariant mass distributions in \verb!plots.ps!.
Note that the NLO prediction is smoother because we have generated 
enhanced weighted events, cf.\ the \verb!Run.NLO.dat! file.

\subsection{Top backgrounds}
Change back to the original directory and then
go to \verb!Backgrounds_TTBar!.
Have a look at the \verb!Run.dat! input file.

Run the setup using
\begin{verbatim}
  <sherpa_prefix>/bin/Sherpa
\end{verbatim}
Once {\sc Sherpa} has finished, plot the results of the analysis using
\begin{verbatim}
  ./plot_results.sh
\end{verbatim}
and have a look at the invariant mass distributions in \verb!plots.ps!.

Now you can devise a strategy to reduce the Standard-Model backgrounds.

\appendix

\section{The history of the MC4BSM series of workshops}

The aim of the MC4BSM series of mini-workshops was to bring together theorists and experimentalists
who are interested in developing and using Monte Carlo tools for BSM Physics 
in preparation for the analysis of data from the LHC. Since simulation of low energy 
supersymmetry (and the MSSM in particular) was already available from several excellent tools, 
the focus of the MC4BSM workshops (as the name suggests) was on simulation tools for alternative 
TeV-scale physics models. The main goals of the workshops were:
\begin{itemize} 
\item To survey what is available and provide user feedback on experiences with Monte Carlo tools for BSM.
\item To identify promising models (or processes) for which the relevant tools are still missing 
and start filling up those gaps.
\item To propose ways to streamline the process of going from from theory models to 
collider events (see Fig.~\ref{fig:path}).  If this process were made more user-friendly,
then a larger community of people would become involved in serious collider studies beyond 
the MSSM.
\end{itemize}

The previous MC4BSM workshops were held at Fermilab (2006), Princeton (2007), CERN (2008),
UC Davis (2009) and Copenhagen (2010). Many new tools 
(e.g.~{\sc Bridge} \cite{Meade:2007js}, {\sc Marmoset} \cite{ArkaniHamed:2007fw}
{\sc Catfish} \cite{Cavaglia:2006uk}, {\sc Pythia-UED} \cite{ElKacimi:2009zj}, etc.)
and model implementations 
(e.g.~the left-right twin Higgs model \cite{Goh:2006wj,Dolle:2007ce}, 
the Littlest Higgs model with $T$-parity \cite{LHTparity},
Minimal UED \cite{Datta:2010us}, etc.)
have been reported at these workshops. The latest, 6th installment in the MC4BSM series 
took place on the beautiful Cornell University campus in Ithaca, NY in March 2012. The weather 
was unseasonably warm and contributed to the lively atmosphere at the workshop.

\acknowledgments
We thank all students who took the tutorial at the workshop for their feedback, support and enthusiasm.
O.~Mattelaer thanks Tim Stelzer for helpful comments on the tutorial.
K.~Matchev and M.~Perelstein thank the Aspen Center for Physics (funded by NSF Grant \#1066293)
for hospitality during the completion of this write-up.
C.~Duhr is supported by the ERC grant ``IterQCD''.
C.~Grojean is partly supported by the European Commission under the ERC Advanced Grant 226371 MassTeV and the contract PITN-GA-2009-237920 UNILHC.
S.~Hoeche's work was supported by the U.S.~Department of Energy under contract DE-AC02-76SF00515, 
and in part by the U.S.~National Science Foundation, grant NSF-PHY-0705682 (The LHC Theory Initiative).
K.~Matchev is supported in part by a U.S.~Department of Energy grant DE-FG02-97ER41029.
O.~Mattelaer is a ``Chercheur post-doctoral logistique FNRS/FRS''  and a fellow of the Belgian American 
Education Foundation. His work is partially supported by the Belgian Federal Office for Scientific, 
Technical and Cultural Affairs through the `Interuniversity Attraction Pole Program - Belgium Science 
Policy' P6/11-P and by the IISN ``MadGraph'' convention 4.4511.10.
M.~Park is supported by the CERN-Korea fellowship through National Research Foundation of Korea.
M.~Perelstein is supported by the U.S.~National Science Foundation through grant PHY-0757868 and CAREER grant PHY-0844667.


\end{document}